\documentclass[preprint, authoryear,12pt]{elsarticle2} 

\usepackage{amssymb}
\usepackage{graphicx}
\usepackage{amsmath}
\usepackage{setspace}
\usepackage{geometry}
\usepackage{endnotes}
\usepackage{multirow}
\usepackage{float}
\usepackage{amsfonts}
\usepackage{threeparttable}
\usepackage{xcolor}
\usepackage{eurosym}
\usepackage{lscape}
\usepackage{enumerate}
\usepackage[english]{babel}

\begin{document}
\begin{frontmatter}

\title{The Composition of Wage Differentials between Migrants and
  Natives}

\author[soton]{Panagiotis Nanos} %
\ead{P.Nanos@soton.ac.uk}

\author[marseille,soton]{Christian Schluter} 
\ead{christian.schluter@univ-amu.fr}  

\address[soton]{Economics Division, University of Southampton, Highfield, Southampton, SO17 1BJ, UK}
\address[marseille]{Aix-Marseille Universit\'e (Aix-Marseille School
  of Economics) , CNRS \& EHESS, Centre de la Vieille Charit\'e,
13002 Marseille, France}


\bigskip 
\begin{abstract}

We consider the role of unobservables, such as differences in search frictions,
reservation wages, and productivities for the explanation of wage
differentials between migrants and natives. We disentangle these by estimating
an empirical general equilibrium search model with on-the-job search
due to \cite{Bontempsetal1999} on segments of the labour market defined
by occupation, age, and nationality using a large scale German
administrative dataset. 

The native-migrant wage differential is then decomposed into several
parts, and we focus especially on the component that we label
``migrant effect'', being the difference in wage offers between natives
and migrants in the same occupation-age segment in firms of the same
productivity. Counterfactual decompositions of wage differentials
allow us to identify and quantify their drivers, thus explaining
within a common framework what is often labelled the unexplained wage
gap.

\end{abstract}

\begin{keyword}
immigrants \sep decomposition of wage differentials \sep job search \sep turnover 

\textit{JEL Classification:} J31 \sep J61 \sep J63
\end{keyword}
\end{frontmatter}

\bibliographystyle{chicago}
\section{Introduction}

The empirical literature on the labour market experience of immigrants
often focuses on differences in observable characteristics between
migrants and natives to explain wage differentials. Less explored is
the role of unobservables, such as differences in search frictions,
reservation wages, and productivities. Yet, it is precisely these
factors that modern search theory emphasises to be important for wage
dispersion. We examine and disentangle the role of these various
unobservables in explaining migrant-native wage differentials by
adapting to the migrant context the empirical general equilibrium
search model with on-the-job search due to \cite{Bontempsetal1999}. 

The estimation of this structural model
on segments of the labour market defined by occupation, age, and
nationality enables us to decompose the native-migrant wage
differential into several parts. In particular, we focus on the
component that we label ``migrant effect'', being the difference in
wage offers between similar native and immigrant workers in firms of the same
productivity. This effect is of interest as we thus
  control for firm-level differences as measured by their
  productivities, which have
  recently been shown using firm-level data to contribute systematically to the wage gap
  (\cite{Aydemir&Skut} in the case of Canada, \cite{Damas} for Portugal, and \cite{BartolucciMigration} for the German
  case).\footnote{The migrant effect corresponds to within-firm wage differentials of workers with similar observable characteristics reported in these papers.} One particular advantage of our approach is that we do not
  require firm-level data (data confidentiality promises usually deny public access), as the
  productivity distribution emerges as an equilibrium relationship. We
estimate the migrant effect on internationally accessible German
administrative data, the scientific use file known as IABS which is a
2\% subsample of the German employment register. 
This enables us to contribute to the recent literature
on the immigrant-native wage gap as follows. While the role of observables is
well understood for explaining the wage gap, the role of unobservables is
less so. Such wage gaps arise when, for instance, migrants have systematically lower
reservation wages (whose role is examined in detail in \cite{Albrecht&Axell}),
or when firms in a migrant-native segmented labour market (which we
discuss below) are less productive in the migrant segment, or when
wage-posting firms in one segment
derive greater monopsony power from e.g. greater search frictions.
Our analysis focuses on the roles of differences in the job
turnover parameters, behavioural differences induced by differences in
reservation wages, and productivity differences.\footnote{ 
The migrant effect is not synonymous with (taste-based) discrimination as we do not
model this explicitly (for two approaches see \cite{Bowlus&Eckstein}, and \cite{FlabbiIER}). Instead, similar to \cite{Bowlus1997} and \cite{BartolucciGender} in the context of the
gender wage gap, we have an indirect link: 
if market discrimination exists and influences behavioural patterns,
it will be captured in those parameters, while other avenues of
wage-impacting discrimination will be picked up by the productivity
distributions. In contrast to costly taste-based
  discrimination, the new monopsony theory suggests the possibility of
  profitable monopsonistic discrimination stemming e.g. from
  differential search friction (\cite{manning2003book}).  For instance \cite{Barth&Dale-Olsen} and \cite{Hirschetal} consider the gender
  wage gap in the light of this. We relate the migrant effect to the \cite{Hirsch&Jahn} analysis of 
monopsonistic discrimination in Section \ref{sec:decomposition}.} Within a common
framework, we establish the relative importance of each of
these factors.  
Having estimated the model's parameters and thus the
actual wage gap and migrant effect, we quantify the 
roles of the various unobservables in several counterfactual experiments. 

The structural model is estimated on a large German administrative panel. Germany is a
particularly interesting and relevant case since it hosts the largest
numbers of foreign nationals in Europe, and immigration is known to be
predominantly low-skilled. According to Eurostat, 7.13 million
foreign nationals resided in Germany in 2010, about 8.7\% of the total population. 
The size of the IABS allows us to stratify the analysis by nationality, occupation
and age. The resulting subsamples are sufficiently large to permit
precise estimation of the model's structural parameters. Moreover,
since this is administrative data, the usual concerns about the
quality of survey data in a migrant context (sample size, measurement
accuracy, and use of retrospective information) are absent. 

We briefly describe some aspects of our applications of the structural model. In order
to control for heterogeneity in observables, we follow common
estimation practice in the search-theory literature by partitioning the labour market into many
segments. These segments are defined in terms of occupation, age, and 
nationality.\footnote{The term ``nationality'' rather than ``immigrant
  status'' is used here for greater precision given the coding practices
  of the German Statistical Office. Most German data sources
  record nationality and not country of
  birth since German nationality was conferred by descent until the year 2000, when Germany changed its legislation to \textit{ius soli} (this change does not affect our sample).} Given the
skill profile of migrants, we consider only the low and medium skill
occupations.
Each segment is thus assumed to be potentially a separate labour market,
characterised by its own job turnover parameters (the job arrival
and separation rates). Turning to the unobservables (for the
econometrician), firms in each segment differ in terms of
productivity, and workers differ in terms of reservation wages. 
Such reservation wage heterogeneity is plausible
given the absence of a legal minimum wage in Germany, and the fact
that the location decisions of labour migrants in Roy-style models
are usually based on comparisons of expected incomes in source and host
country. Migrants might trade-off wage and non-wage job
  characteristics differently to natives, given their well-known clustering.
Besides this preference component, reservation wages also
feature an institutional one, but this is less important
as contributory unemployment
insurance benefits are independent of immigrant status.

The assumption of separate markets for natives and
immigrants and the associated notion of job segmentation conforms to
existing international empirical evidence. For instance, using
Portuguese data, \cite{Damas} shows that immigrants ``work in
different industries and occupations than natives'' (p.10), and the
sorting of immigrants is also observed by \cite{Aydemir&Skut}
for Canada. As regards Germany, \cite{DAmurietal} observe that recent immigrants are significantly more likely
to compete with established immigrants rather than with natives.
\cite{Velling} is an early paper to
report ``evidence of strong occupational segregation'' (p.1) between
natives and immigrants. This finding has recently been reaffirmed by
\cite{Lehmer&Ludsteck}, \cite{Bruecker&Jahn}, \cite{BartolucciMigration}, and \cite{Glitz2012} who concludes that ``ethnic segregation [..]
is endemic in the German labour market'' (p.15).\footnote{At the same time,
    these papers provide complementary perspectives on the
    native-immigrant wage gap in Germany: descriptive Oaxaca-Blinder
    decompositions (\cite{Velling}, \cite{Lehmer&Ludsteck}), wage
    setting (\cite{Bruecker&Jahn}), monopsonistic discrimination
    (\cite{Hirsch&Jahn}), while \cite{BartolucciMigration} provides an
    interpretation in terms of taste-based discrimination.  
 \cite{DAmurietal} pursue a different concern and
 estimate the wage and employment effects of recent immigration in
 Western Germany (and find little evidence for adverse effects on
 native wages and employment levels).}
This segmentation is also
consistent with the evidence of strong occupational immobility
we find in our data (which has also been observed for other countries,
e.g. by \cite{Damas} for Portugal).\footnote{The segmentation
  assumption has also been imposed routinely in recent search-based structural
  analyses of the gender wage gap. For instance, \cite{FlabbiIER}
  considers only whites possessing a college degree, \cite{Bowlus1997}
  considers two education groups, and \cite{BartolucciGender} considers
  four sectors and two skill groups. Our partition is finer as we also
  consider three age groups in addition to our three occupation groups
  (and our estimates remain unbiased should the true partition be such
  that some segments be aggregated).}

For each occupation-age segment, we estimate using maximum likelihood
the job turnover parameters, the parameters characterising the reservation wage
distribution, and the firms' productivity distribution. 
We find substantial differences in Germany
between natives and foreigners. 
The segment-specific raw average log wage gaps in our
  data range from .09 to .45, the overall log wage
    gap being .22, which is in line with reports in the literature for
  Germany (e.g. \cite{DustmannetalEER} report an unconditional average
  log wage gap of .23, \cite{Hirsch&Jahn} report a gap of .2,
  while \cite{Lehmer&Ludsteck} report predicted wage gaps ranging
  from .08 to .44 depending on nationality). Turning to the
  qualitative implications of our model estimates, we find that
migrants experience job separations more often than natives
but also find jobs more quickly. However, the net effect
  is such that migrants typically experience greater search frictions.
The job turnover parameters decline
in age. Across all segments and nationality, transitions into new
jobs happen more quickly than transitions into unemployment.
This finding of migrants' higher job separation and offer
  rates is consistent with differences in employment protection; in
  particular, \cite{Sa2011} reports that
  migrants in Germany are much more likely than natives to work on
  temporary contracts.
As regards the reservation wage distribution, there are some workers
in all segments with high reservation
wages who turn down new job offers when wage offers are too
low. However, migrant workers are less demanding on average
than natives.

Migrants receive wage offers that are lower than those for natives
controlling for the same productivity. This migrant effect is the largest for
clerks and service workers, and small for unskilled workers.
In particular, the average migrant effect for the skilled ranges between 12\%
and 15\% of the average wage gap, and for clerks and service
workers the range is 23\% to 39\%. For all occupation groups, the migrant effect
declines across age groups. These estimates imply that the largest
part of the within-group native-migrant wage gap is explained by differences in the
productivity distribution (one explanation for such
  productivity differences is advanced in \cite{Damas}). At the same time, the migrant effect is
significant in many segments, and, if expressed in terms of the average segment-specific wage of natives, it is found to be consistent with
estimates of ``unexplained wage differences'' reported in the
literature for Germany based on standard Oaxaca-Blinder decompositions
(for instance, \cite{Lehmer&Ludsteck} report a range from 4 to
17\%) or complementary approaches (\cite{Hirsch&Jahn} report 6\%
while \cite{BartolucciMigration} suggests discrimination effects ranging
between 7 and 17\%). Our counterfactual decomposition approach 
allows us to quantify the (marginal and joint) roles of the underlying
drivers of the migrant effect in terms of labour market turnover
parameters and behavioural differences captured by the reservation wage
distribution. We find that reducing the job separation rate for
migrants to that of natives typically leads to a large reduction
in the migrant effect. This is of interest to policy makers since this
parameter is targetable by e.g. deploying measures to improve
migrants' employment protection.

This paper is organised as follows. In Section \ref{model_section}, we set out the model
as well as the estimation approach. A validation exercise, reported in
the Appendix, verifies that the estimation of
the structural parameters works well. Section \ref{sec:decomposition}
introduces the migrant effect, the decomposition of the actual wage
differential, and the counterfactual scenarios in the context of the simulated
data (which are later re-examined in Section \ref{sec:experiments}
with the real data). Section \ref{data_section} describes the data used for
the analysis. The estimation results are presented in Section
\ref{results_section}, and the resulting decompositions in Section
\ref{sec:experiments}. Section \ref{conclusion} concludes.

\section{The Analytical Framework}\label{model_section}

The search model with wage-posting and on-the-job search has been described
and discussed extensively before in the literature. Therefore, only its most
salient features will be outlined. We use the extension of the \cite{Burdett&Mortensen} model, and the subsequent empirical generalisation and
implementation of \cite{VanDenBerg&Ridder}, due to \cite{Bontempsetal1999}. This extends the basic setting by introducing productivity
heterogeneity among firms, which improves the fit of the model to wage data,
and heterogeneity among workers in terms of the
unobserved opportunity cost of employment, which improves the fit to
the unemployment duration data. As discussed above, the
  latter is very plausible in the migration context against the
  background of Germany's institutional rules.

The labour market is partitioned into many segments, defined in our
empirical implementation by age, occupation and nationality. Each
segment is considered as a labour market for which the following model
and estimation approach applies. The structural parameters are of
course allowed to vary across segments, but for notational simplicity
we suppress a segment index. This segmentation assumption precludes
individuals moving from one segment to another, which is consistent
with the evidence of occupational immobility in Germany presented
below and the external evidence discussed in the Introduction.
If the labour market is integrated over some
stipulated segments, then the estimates of the structural parameters
should be the same statistically; the segments can then be added to
improve estimation efficiency. In line with the segmentation
hypothesis we find that the estimated
structural parameters differ across occupation-age-nationality
groups. We proceed to outline the model for one labour market segment.

\subsection{The Model of a Labour Market Segment}

The labour market segment is populated by a fixed continuum of workers with
measure $M$, and a fixed continuum of firms with measure normalised to one. Firms
differ in terms of (the marginal) productivity (of labour) $p$ with
distribution $\Gamma$. Unemployed workers differ in terms of their
reservation wages $b$ with distribution $H$.

At any point in time, a worker is either unemployed or employed, and
searches for jobs both off and on the job. Individuals draw offers by
sampling firms using a uniform sampling scheme. Jobs are terminated at the
exogenous rate $\delta $, and job offers arrive at the common rate $\lambda $
irrespective of the worker's state. This is a restrictive assumption
but necessary for identification.\footnote{ This
    assumption yields, for the unemployed, a simple solution for the
    opportunity cost of employment: it is simply equal to $b$. If job
    offer arrival rates were to differ, \cite{Mortensen&Neumann}
    show that this opportunity cost would be an intractable function
    of all the primitives of the model, leading to feedback to
    workers' optimal strategies from wages and firm behaviour.
} Let  $k=\lambda /\delta $.

Job offers are, of course, unobservable to the
econometrician. The job offer distribution is denoted by $F$, whereas the
observable wage or earnings distribution (i.e. of accepted wages) is denoted
by $G$. Let  $\left[ \underline{w},\overline{w}\right] $
denote the support of $F$, and, 
for notational convenience, $\overline{F} =\left[
1-F \right] $. $F$ is related to $G$ through an equilibrium condition implied by
the theoretical structure. Firms post wages and there is no bargaining.%
\footnote{For an analysis of wage determination in the presence of
  heterogeneity, search on-the-job, and strategic wage bargaining, see
  \cite{Cahucetal2006}. They find no significant bargaining power for
  intermediate and low skilled workers in France.}

Workers are risk neutral and maximise their expected steady state discounted
future income. Their optimal strategy has the reservation wage property: an
employed individual moves to a new employer if the offered wage exceeds the
current wage (so the model does not allow for wage cuts); an unemployed
individual accepts a new job if the offer exceeds $b$, and otherwise rejects
the offer and remains unemployed. On-the-job search thus generates
further ex-post heterogeneity in reservation wages.

In steady-state equilibrium, the flows of workers into and out
of the unemployment pool are equal, which determines the unemployment rate $u
$. Consider the stock of employed workers who earn a wage less than or
equal to $w$. 
Two sources constitute the outflow from this stock, namely: (i) exogenous job separations at rate $\delta $ and
subsequent transits into unemployment, and (ii) wage upgrading as employed
workers move to poaching firms. The combined outflow is thus
$(1-u)G(w)(\delta + \lambda \overline{F}(w) )$.
The flow into this stock consists of unemployed individuals
who receive wage offers above their reservation wage. Conditional on
$b$, the probability of this event is $u \lambda [F(w) - F(b)]$. The marginal inflow is
obtained by integrating up to $w$ over the distribution of $b$ in the
stock of the unemployed. Denoting the latter by $H_u$, the steady
state equation for the labour market yields the relationship between
$H_u$ and $H$, namely $uH_u (b) = \int _{-\infty} ^b [ 1+k
\overline{F} (x)]^{-1} dH(x)$.

Equating inflows and outflows relates
the wage offer distribution $F$ to the realised wage distribution 
$G$. To be precise, \citet[Proposition~2]{Bontempsetal1999}
show that the unemployment rate
$u$ and the actual wage distribution $G$ satisfy
\begin{equation}
u = \left[ \frac{1}{1+k}H\left( \underline{w}\right) +\int_{\underline{w}}^{%
\overline{w}}\frac{1}{1+k \overline{F}\left( x\right) }dH\left(
x\right) \right] +\left[ 1-H\left( \overline{w}\right)
\right]  \label{u_IER_1999}
\end{equation}
\begin{equation}
G\left( w\right) =\frac{H\left( w\right) -\left[ 1+k \overline{F}\left(
w\right) \right] \left[ \frac{1}{1+k}H\left( \underline{w}\right)
+\int_{\underline{w}}^{w}\frac{1}{1+k \overline{F}\left( x\right)
}dH\left( x\right) \right] }{\left[ 1+k \overline{F}\left( w\right)
\right] \left( 1-u\right) } . \label{G_IER_1999}
\end{equation}%

Risk neutral firms have constant-returns-to-scale technologies, and post
wages that maximise steady state profit flows, the profit per worker being $%
p-w$. Firms do not observe the reservation wage of a potential employee. In
equilibrium, firms offer wages to workers that are smaller than their
productivity level, so firms have some monopsony power. \citet[Proposition 9]{Bontempsetal1999} show that in equilibrium there exists an increasing function $K$
which maps the productivity distribution $\Gamma$ into the wage offer
distribution $F$, so that the wage offer satisfies $w=K(p)$ with 
\begin{equation}
K\left( p\right) =p-\left[ \frac{\underline{p}-\underline{w}}{\left(
1+k\right) ^{2}}H\left( \underline{w}\right) +\int_{\underline{p}}^{p}\frac{%
H\left( K\left( x\right) \right) }{1+k\left[ 1-\Gamma \left( x\right) \right]^2
}dx\right] \frac{\left[ 1+k\left[ 1-\Gamma \left( p\right) \right] \right] ^2}{%
H\left( K\left( p\right) \right) }  \label{K_IER_1999}
\end{equation}%
and $F\left( w\right) =\Gamma \left( K^{-1}\left( w\right)
\right)$. Hence given the frictional parameter $k$, the reservation
wage distribution $H$ and the productivity distribution $\Gamma$,
equation (\ref{K_IER_1999}) yields the wage offer distribution $F$, which then
via (\ref{u_IER_1999}) yields the equilibrium unemployment rate and
through (\ref{G_IER_1999}) the actual wage distribution $G$. 

Our dataset does not include measures of firm productivity but, of course,
extensive wage data. Using expressions of the key
quantities in terms of the actual wage density $g$, the productivity
distribution $\Gamma $ becomes estimable. In particular, it can
be shown that 
\begin{equation}
\left( 1-u\right) = \frac{k}{\left( 1+k\right) \int_{\underline{w}}^{%
\overline{w}}\frac{g\left( t\right) }{H\left( t\right)
}dt},  \label{SS_A2}
\end{equation}
\begin{eqnarray}
\frac{1}{\left[ 1+k\overline{F}\left( w\right) \right] } &=&\left( 1-u\right)
\int_{\underline{w}}^{w}\frac{g\left( t\right) }{H\left( t\right) }dt+\frac{1%
}{\left[ 1+k\right] }.  \label{SS_A1}
\end{eqnarray}%
Equation (\ref{SS_A2}) follows from
(\ref{SS_A1}) with $w=\overline{w}$. The equilibrium
productivity levels are
\begin{equation}
p=K^{-1}\left( w\right) =w+\frac{H\left( w\right) }{2\left( 1-u\right) g\left(
w\right) \left[ 1+k\overline{F}\left( w\right) \right] +h\left( w\right) }.
\label{Prod_A}
\end{equation}

\subsection{Identification}

We seek to estimate this model using data by labour market
segment on employment and unemployment durations, as well as data on
wages and accepted wage offers. These data are sufficient to
identify\footnote{\cite{Eckstein&VanDenBerg} discuss identification issues in
empirical search models more generally.}
the structural parameters, once the reservation wage distribution is
parametrised.  We assume that $H$ is a normal distribution with
unknown location and scale parameters, $(\mu,\sigma) \equiv
\theta$. Since arrivals of job offers and separations are assumed
to follow Poisson processes, sojourn times are exponentially distributed.

In particular, the wage data identify the wage distribution $G$, and
the minimum and the maximum of the observed wages identify
the infimum $\underline{w}$ and the supremum $\overline{w}$
of the wage offer distribution. The steady state flow equations in
form of (\ref{SS_A2}) and (\ref{SS_A1}) then identify the wage offer
distribution $F$ given $\lambda/\delta$ and $H(.;\theta)$, which yield
the productivity distribution $\Gamma$ via (\ref{K_IER_1999}).
The job separation rate is identified from job durations ending in a
transition to unemployment, as these are exponential variates with
parameter $\delta$, the mean duration being $\delta^{-1}$. Job durations ending in a transition to another job
with wage $w$ are exponential with parameter $\lambda
\bar{F}(w)$. Together with unemployment durations ending in a
transition to a job with wage $w$ these identify the remaining
parameters $\lambda$ and $\theta$. Since the reservation wage is
unobservable, the marginal unemployment durations are mixtures of
exponentials, $\Pr\{T_u \leq t|b \leq w\}=1- \int
  _{-\infty} ^w \exp(-\lambda \bar{F}(b)t) dH_u (b;\theta|b
\leq w)$. 

Absent such mixing, when $H$ is degenerate and all
agents accept all wage offers above the common reservation wage, transitions to a new job from each
labour market state would permit separate identification of the job offer
arrival rates, and thus would give rise to testable overidentification restrictions.
In the presence of unobservable heterogeneity captured
by $H$, overidentification
restrictions only arise with additional data that would permit, for
instance, an independent estimation of the wage offer distribution
(see e.g. \cite{Christensenetal} for such an approach). 

\subsection{Maximum Likelihood Contributions for Labour Market Segments}

The preceding constructive identification argument suggests that we
can estimate the structural parameters using maximum likelihood on our
data on unemployment and employment durations and wages. 
The likelihood contributions we consider in detail next differ
slightly from those in \cite{Bontempsetal1999} since our data are
flow and not stock samples. The validation exercise reported in \ref{sec:validation}
verifies the good performance of our estimation procedure on
artificial data. The density of accepted wages, and thus $G$, is estimated using kernel
methods, and enters all likelihoods as a nuisance parameter.

Consider first the likelihood contributions of unemployed
agents. Since the unemployment rate is a function of the model
parameters, it needs to enter the sampling plan.
In equilibrium, the probability of encountering an unemployed
individual is given by (\ref{SS_A2}). Since the reservation
wage $b$ is unobservable, it needs to be integrated out. We
distinguish between individuals for whom $b \leq \underline{w}$ as
they accept all job offers, a mass of $H(\underline{w})$, and those
for whom $b > \underline{w}$ as they reject offers below $b$. 
Recall that $F\left( \underline{w}\right) =0$, and we assume that all
individuals included in our sample would accept at least one wage
offer $w\in \lbrack \underline{w},\overline{w}]$. This implies that
the $\sup $ of 
$H$ is lower than the $\sup $ of $F$, $\overline{b}\leq \overline{w}$, so
this specification does not take into account cases of permanently
unemployed individuals. Conditional on $b$, the distribution of unemployment durations
in our flow sample is exponential with parameter $\lambda \overline{F}\left( b\right)$. 
The accepted wage, $w$, is a realisation of the wage offer
distribution truncated at $b$: $f\left(w\right) / \overline{F}\left(
  b\right)$. The  likelihood contribution of an unemployed $L_{u}$ is
thus, having substituted out $u$,
\[
L_{u}\left( \lambda,\delta ,\theta \right) =\lambda ^{\left( 1-d_{r}\right) }\exp \left(
-\lambda t\right) \frac{H\left( \underline{w}\right) }{1+k}\left[ f\left(
w\right) \right] ^{\left( 1-d_{r}\right) }+ 
\]
\begin{equation}
+\int_{\underline{w}}^{w}\left\{ \left[ \lambda \overline{F}\left( b\right) %
\right] ^{\left( 1-d_{r}\right) }\exp [-\lambda \overline{F}\left( b\right)
t]\left[ \frac{f\left( w\right) }{\overline{F}\left( b\right) }\right]
^{\left( 1-d_{r}\right) }\frac{1 }{\left[ 1+k\overline{F}%
\left( b\right) \right] }\right\} dH(b),  \label{Lu_A}
\end{equation}%
where $d_{r}$ is a dummy variable equal to one if
the spell is right-censored (the only relevant censoring in our
data). In this case it is only known that the unemployment
duration exceeds $t$.

We turn to the likelihood contributions of employed workers, denoted
by $L_{e}$. The probability of sampling an employed individual
receiving a wage $w$ is $\left( 1-u\right) g\left( w\right) $. We have
further data on the job duration and the exit state. Let $v$ be a dummy
variable equal to one if the destination of an employment spell is
unemployment, and zero if the destination is another job.
We have two competing risks: Exits
to unemployment occur with probability $\delta / \left[ \delta +\lambda
  \overline{F}\left( w\right) \right]$ and exits to higher paying jobs occur with
probability $\lambda \overline{F}\left(w\right) / \left[\delta
+\lambda \overline{F}\left( w\right) \right]$.
Conditional on being employed with
wage $w$, the job duration has an exponential distribution with
parameter $\left[ \delta +\lambda \overline{F}\left(w\right) \right]$.
If a transit to unemployment is observed at duration $t$, this implies
that the duration of the other latent risk factor exceeds $t$, the
joint density factorises, and we have $\delta \exp(-\delta t)
\exp(-\lambda \overline{F}\left(w\right))$.
Therefore
\begin{equation}
L_{e}\left( \lambda,\delta ,\theta \right) =\left( 1-u\right) g\left( w\right) \exp
\left\{ -\left[ \delta +\lambda \overline{F}\left( w\right) \right]
  t\right\}  \times \left\{ \delta ^{v}\left[ \lambda \overline{F}\left( w\right) %
\right] ^{\left( 1-v\right) }\right\} ^{\left( 1-d_{r}\right) }, \label{Le_A}
\end{equation}%
where $(1-u)$ is given by equation (\ref{SS_A2}). If an employment spell is
right-censored, indicated by $d_r$, we only know that the job
duration exceeds $t$.

\subsection{Migrants, Natives, Wage Differentials and the Migrant
  Effect: Concepts and Simulated Data} \label{sec:decomposition}

We develop an illustrative example in order to introduce our key concepts.
Consider two labour market segments, one occupied by natives (N) and the other
by immigrants (F). Workers in either segment exhibit the same observable
characteristics (in our empirical application below we consider the
same skill and age group). We calibrate the two segments 
(in line with the empirical results) as follows: the job
turnover parameters of migrants are assumed to be higher than those of natives,
$\delta_F=.016 > .005= \delta_N$ and $\lambda_F=.13 > .07=\lambda_N$,
while natives have higher mean reservation wages, $\mu_F = 45 < 60=
\mu_N$. The productivity distribution in the
segment for natives is assumed to first order stochastically
dominates that of migrants: $\Gamma_F
(p)=1-(\underline{p}_F/p)^{\alpha}$ and $\Gamma_N
(p)=1-(\underline{p}_N/p)^{\alpha}$ with $\alpha=2.1$,
$\underline{p}_F=40$, and $\underline{p}_N=50$. 
The validation exercise reported in \ref{sec:validation} discusses the
estimation results.

 \begin{figure}[htbp]
\centering
\caption{Wage offer curves for natives and migrants, and the ``migrant effect''.}
\includegraphics[scale=0.3] {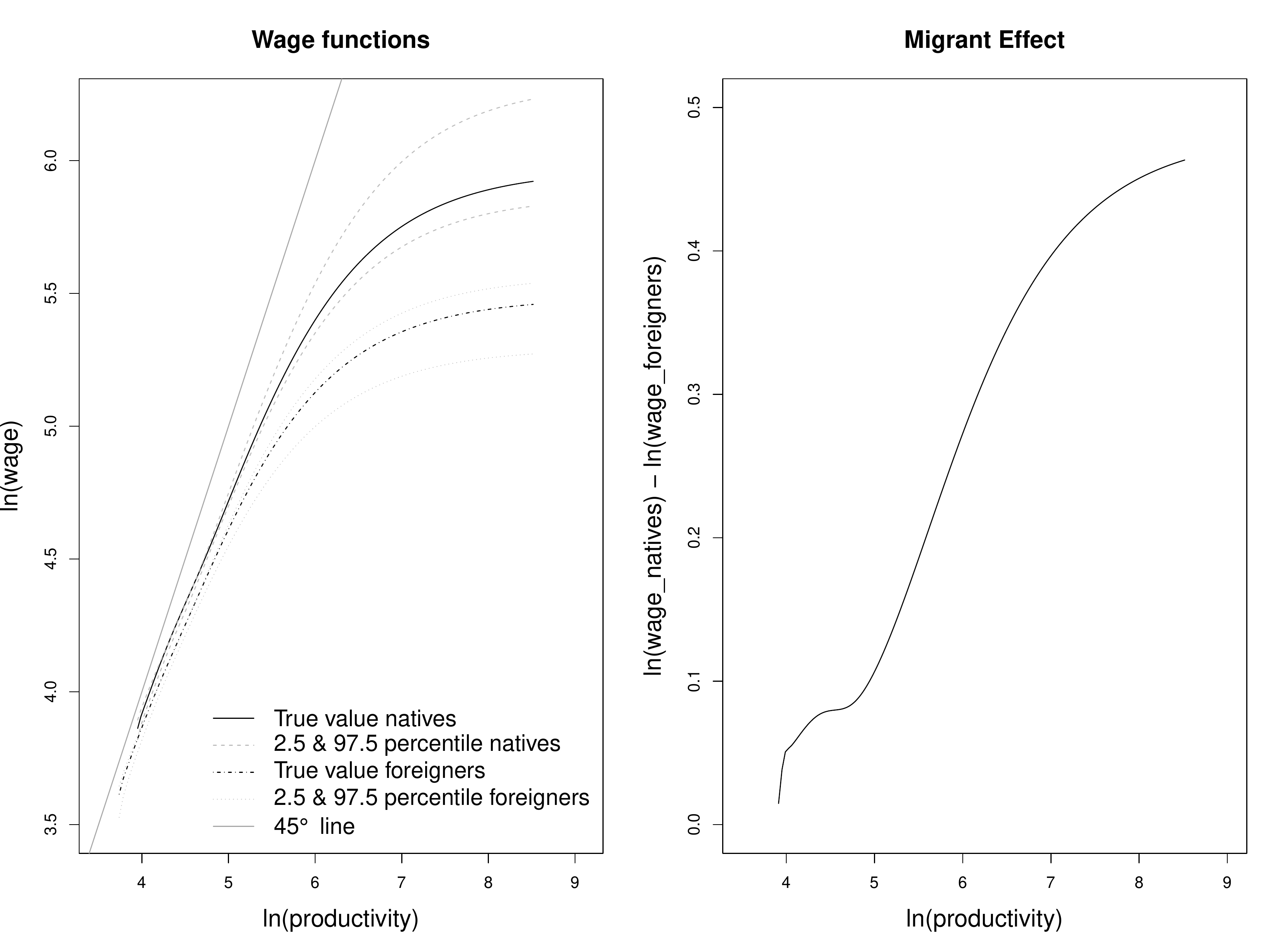}
\label{fig:Migrant_effect}
\end{figure}

For this economy, the aggregate wage gap is substantial (equal to
32.02), but differences in the productivity distributions are likely
to play an important role (recall the discussion in the Introduction). 
Figure \ref{fig:Migrant_effect} Panel A depicts the
resulting wage offers given by (\ref{K_IER_1999}) as functions of
productivity. These enable us to consider a component of the wage gap
which we label ``migrant effect'', depicted in Panel B, being the difference in wage offers
between similar native and immigrant workers in firms of the same
productivity: $ w_N(p)- w_F(p)$. This effect is of interest since we thus control for
firm-level differences as measured by their productivities. 

This concept of the migrant
effect suggests to decompose the aggregate wage
differential\footnote{For a decomposition of wage
differentials in a reduced form
 setting, see \cite{Dustmann&Theo}. Note that their
  decomposition considers, as we do, the wage offer function, but
  their empirical approach does not recover it from the data.} 
between migrants and natives, 
$\int_A w_N(p)d\Gamma_N (p) - \int_A w_F(p)d\Gamma_F (p)$,
into the aggregate migrant effect and a
weighted difference between firm productivities (where $A$ denotes the
intersection of the supports of the productivity distributions).
Solving for the aggregate migrant effect, we thus have
\begin{eqnarray}
\int_A \left[w_N(p) - w_F(p) \right]d\Gamma_N (p) &=&
\int_A w_N(p)d\Gamma_N (p) - \int_A w_F(p)d\Gamma_F (p) 
 \label{decomp1} \\
&-& \int_A w_F(p)d\left[\Gamma_N (p) - \Gamma_F (p) \right]. \notag
\end{eqnarray}

We briefly comment on the
  relationship between the migrant effect and the concept of monopsonistic
  discrimination, as examined in e.g. \cite{Hirsch&Jahn}. The
  latter is measured by these authors indirectly from a search-model
  inspired decomposition of the long run wage elasticity of labour
  supply using reduced-form job separation models that are estimated separately on
  data for migrants and natives. In our model, greater monopsony
  power of firms (measured by the absolute or relative distance
  between productivity, i.e. the 45 degree line, and wages as illustrated in Figure 1.A) in the
  migrant segment gives rise to the migrant effect. Our approach
  enables us to go beyond measuring the migrant effect, as we
  explain it within a common framework in terms of the relative
  importance of differences in the
  job turnover parameters and behavioural differences induced by
  differences in reservation wages. 
 In particular, a closer inspection of (\ref{K_IER_1999}) shows that
 the wage offers are complicated functions of these structural
 parameters, $w_i(p|\underline{p}_{i},\alpha_{i},\mu_{i},\sigma_{i},\lambda_{i}
,\delta_{i})$ for $i \in\{N,F\}$.

\subsubsection{Counterfactual Wage Decompositions}\label{sec:counterfactual.sim}

In order to identify the principal drivers of the migrant effect, and 
to conduct policy experiments, we  consider next a second
decomposition of the wage gap based on counterfactuals. In particular, we
ask: what would be the migrant effect and the wage differential if
one group is imputed counterfactually parameter values of the other
group? For instance, choosing natives as the reference group and
equalising counterfactually the reservation wage distribution parameters $(\mu,
\sigma)$, the counterfactual migrant effect is, using (\ref{decomp1}),
\begin{eqnarray}
&& \int_{A}[w_N(p|\underline{p}_{N},\alpha_{N},\mu_{N},\sigma_{N}
,\lambda_{N},\delta_{N})-w_F(p|\underline{p}_{F},\alpha_{F},\mu_{N},\sigma
_{N},\lambda_{F},\delta_{F})]d\Gamma_N(p) \label{counterdecomp1} \\
&=& 
\int_{A}w_N(p|\underline{p}_{N},\alpha_{N},\mu_{N},\sigma_{N}
,\lambda_{N},\delta_{N})d \Gamma_N (p)
-\int_{A}
w_F(p|\underline{p}_{F},\alpha_{F},\mu_{N},\sigma_{N},\lambda_{F},\delta
_{F})d \Gamma_F (p)  \notag \\
&-& \int_{A}w_F(p|\underline{p}_{F},\alpha_{F},\mu_{N},\sigma_{N}
,\lambda_{F},\delta_{F})d[ \Gamma_N (p)- \Gamma_F (p)]  \notag
\end{eqnarray}
with $\Gamma_i (p)=\Gamma_i (p|\underline{p}_{i},\alpha_{i})$ for $i
\in \{N,F\}$.

\begin{table}
\begin{center}
\begin{threeparttable}
\caption{Counterfactual decompositions of the wage differential using
  natives as the reference group.} \label{table:Decomp_SIM}
\begin{tabular}{rllcc} \hline \hline                                
& \multicolumn{1}{l}{Counterfactually } & \multicolumn{1}{l}{Remaining} & \multicolumn{1}{l}{Wage}    & \multicolumn{1}{l}{Migrant}\\
& \multicolumn{1}{l}{equalised para.} & \multicolumn{1}{l}{differing para.}
& \multicolumn{1}{l}{differential}    & \multicolumn{1}{l}{effect}\\
 \hline                               
(1)& & $\underline{p},\alpha,\mu,\sigma,\lambda,\delta$ & 32.022& 6.825	\\
(2)& $\mu,\sigma$ & $\underline{p},\alpha,\lambda,\delta$& 30.096& 3.747	\\
(3)& $\delta$ &$\underline{p},\alpha,\mu,\sigma,\lambda$ & 28.973& 1.954	\\
(4)& $\lambda$ & $\underline{p},\alpha,\mu,\sigma,\delta$ & 34.029&10.032	\\
(5)& $\mu,\sigma,\delta$ & $\underline{p},\alpha,\lambda$ & 27.423&-0.524	\\
(6)& $\alpha,\mu,\lambda$ & $\underline{p},\alpha,\delta$ & 31.694& 6.300	\\
(7)& $\lambda,\delta$ & $\underline{p},\alpha,\mu,\sigma$ & 30.459& 4.328	\\
(8)& $\mu,\sigma,\lambda,\delta$ & $\underline{p},\alpha$ & 28.758 & 1.610	\\ 
(9)& $\underline{p},\alpha$ & $\mu,\sigma,\lambda,\delta$ &	& 4.904	\\
(10)& $\underline{p},\alpha,\mu,\sigma$ & $\lambda,\delta$ &	& 1.932	\\
(11)& $\underline{p},\alpha,\delta$ & $\mu,\sigma,\lambda$ &	& 0.750	\\
(12)& $\underline{p},\alpha,\lambda$ & $\mu,\sigma,\delta$ &	& 7.814	\\
(13)& $\underline{p},\alpha,\mu,\sigma,\delta$ & $\lambda$ &	&-1.842	\\
(14)& $\underline{p},\alpha,\mu,\sigma,\lambda$ & $\delta$  &	& 4.400	\\
(15)& $\underline{p},\alpha, \lambda,\delta$ & $\mu, \sigma$  &	& 2.741	\\
\hline\hline
\end{tabular}
\footnotesize Notes: Based on the DGP given in Appendix Table
  \ref{table:Simulation_Natives}, and the decomposition of equation (\ref{counterdecomp1}). Rows 9$+$: the wage differential
  equals the migrant effect because the productivity distributions are
  the same. 
\end{threeparttable}
\end{center}
\end{table}

Table \ref{table:Decomp_SIM} collects the exhaustive list of possible
counterfactual experiments, and the resulting quantifications of both
the counterfactual migrant effect and wage differential (the first
term on the right hand-side of (\ref{counterdecomp1})). The reference group
consists of natives. In column 1 we list the parameters we
counterfactually equalise, so $(\mu,\sigma)$ in row and experiment 2
is a shorthand for $\mu_F=\mu_N$ and $\sigma_F = \sigma_N$. The
residual parameters enumerated in column 2 constitute thus the sources
of the remaining wage differences. In the first experiment, reported
in row 1, no parameters are equalised, hence the reported results are
based on actual wages (i.e. we use the actual wage decomposition
(\ref{decomp1})). In experiment 9 and later, we equalise the two
parameters of the productivity distribution, $\underline{p}$ and
$\alpha$ (\cite{BartolucciGender} labels such differences in the
productivity distribution parameters ``segregation''). This nils the last term in equation (\ref{counterdecomp1}),
so migrant effect and wage differential are equalised.
In all experiments we use simulated data based on the DGP of Appendix Table
\ref{table:Simulation_Natives} but the results reported next are in
line with our data-based empirical results for the comparative statics
and policy experiments reported in Section \ref{subsec:experiments}.

The actual migrant effect of  6.8, reported in experiment 1, is
substantial, about 21\% of the wage differential. At the same time
this implies that the largest contribution to the native-migrant wage
gap is made by the differences between the productivity
distributions. Turning to the drivers of the migrant effect,
experiments 13-15
consider the marginal roles of $\delta$, $\lambda$, and $(\mu,
\sigma)$. Recalling that $\lambda_F > \lambda_N$ explains the negative
sign in experiment 13. Also note that $\delta_F > \delta_N$, and
$\mu_F < \mu_N$ while $\sigma_F= \sigma_N$. Experiment 14 suggests
that the difference in the separation rates plays a large quantitative
role in the determination of the migrant effect, the latter being
4.4; the complementary insight is that, by experiment 3, equalising
the job separation rates reduces the migrant effect to 29\% of its
former size. 
The differences in mean reservation wages, considered in
experiment 15, leads to a smaller migrant effect of 2.7. The joint
effect of $\delta$ and $(\mu,\sigma)$, reported in experiment 12, equals 7.8,
and is slightly larger than the sum of the two marginal effects.
We defer discussing the policy implications of these
  results to Section \ref{subsec:experiments} as these are similar to
  those based on our empirical results.

\section{The Data}\label{data_section}
The empirical analysis is based on the 2\% subsample of the German
employment register provided by the Institute of Employment Research, known
as IABS (75-04 distribution). For a detailed description of the
dataset, see \cite{Benderetal}. This large administrative dataset for Germany, covering the period
1975-2004 consists of mandatory notifications made by employers to
social security agencies. These notifications are made on behalf of workers,
employees, and trainees who pay social security contributions. This means
that self-employed individuals, civil servants, and workers in marginal
employment are not included. Notifications are made at the beginning and at
the end of an employment or unemployment spell. Information on individuals
not experiencing transitions during a calendar year is updated by means of
an annual report. Hence, we are able to use a flow sample in
our empirical analysis.

Apart from wages, transfer payments, and spell markers, the dataset contains
some standard demographic measures, including nationality, as well as
occupation and firm markers. The education variable is not used since
its problems, particularly in the migrant context, are well-known and
skills are better measured by the occupation (see \cite{Fitzetal} for a detailed
discussion; we do not use the suggested imputations since the
education variable for migrants, when observed, is likely to be of poor quality, as
discussed in \citet[p. 296 point (ix)]{Bruecker&Jahn} and \citet[p.~900]{Lehmer&Ludsteck}).
Wage records in the IABS are top coded at the social
security contribution ceiling. However, this ceiling is not binding
for our population of interest, namely individuals (natives and
foreigners) in low and middle skill occupations. We use real wages in
1995 prices. The occupational information is provided in extensive (three digit codes) but
non-standard form. We therefore map this coding into 10 major groups
based on the International Standard Classification of Occupations
(ISCO-88). The Data Appendix provides some details. Since immigration
is known to be predominantly low skilled, we select from these 10
groups 3 low and middle skilled occupations, namely (1) unskilled
blue-collar workers, (2) clerks and low-service workers, and (3)
skilled blue-collar workers.

The data allows us to distinguish between three labour market states:
employed, recipient of transfer payments (i.e. unemployment benefits,
unemployment assistance and income maintenance during participation in
training programs) and out of sample. Unfortunately, none of the two last
categories corresponds exactly to the economic concept of unemployment. This
issue is discussed in several studies, see e.g. \cite{Fitz&Wilke}. For example, participants in a training program are transfer payment
recipients despite being in employment (they are considered unemployed from
an administrative point of view), while individuals that are registered
unemployed but are no longer entitled to receive benefits appear to be out
of the labour force. Therefore, the dataset provides a representative sample
of those employed and covered by the social security system, but somewhat
mis-represents those in the state of unemployment. For our purposes, all
individuals who are out of sample between two different spells are
classified as unemployed, so only two labour market states are considered:
unemployment and employment. The definition of unemployment used in our
analysis is therefore somewhat broad: we assume that unemployment is proxied
by non-employment, strictly speaking non-employment is an upper-bound for
unemployment.

Nationality is included as a binary variable indicating whether an
individual is German or a foreign national. German nationality is usually
conferred by descent, and not by place of birth. The data set does not
report place of birth. Given this coding practice, some young foreign
nationals might be born and raised in Germany. At the same time,
ethnic Germans who immigrated from the former Soviet Union after the
fall of the Berlin Wall will be classified as German, although they
usually speak little German and have low skills. However, \cite{DustmannetalEER} have
argued that the former issue is ignorable, and we address the second by
repeating the estimation using the subsample of individuals that were
present in the data before the fall of the Berlin Wall, see the analysis
in Section \ref{ethnic_Germans}.

\subsection{The Sample}

The data used in our empirical analysis is restricted to male full-time
workers aged 25 to 55 years old residing in West-Germany (East Germany is
excluded because of the peculiar transition processes taking place in the
wake of unification). This sample is grouped into cells by occupation,
nationality, and age. We define three age groups (25-30, 30-40, and 40-55) to
proxy for potential experience. 
The aim of the grouping is to arrive
at cells in which individuals are fairly homogeneous, and which are
sufficiently large for the subsequent econometric investigation.

\begin{table}[htbp]
\begin{center}
\begin{threeparttable}
\caption{Occupational Immobility: Share of Stayers by Segment} \label{table:occup_im}
\begin{tabular}{cl|cc|cc}
\hline \hline
		&	Age group		 & \multicolumn{2}{c}{Natives}  &\multicolumn{2}{|c}{Foreigners}\\
\hline
Unskilled       &			 & 89.52\%     	&		&88.27\%&		\\
		& Twenties   &      	&  85.72\%	&	&85.45\%	\\
		&Thirties   &      	&  88.03\%     	&	&88.56\%	\\
		& Fourtyplus &      	&  92.54\% 	&     	&92.38\% 	\\
Clerks  	& 			 & 90.06\%      &  		&88.52\%&         	\\	
		& Twenties   &      	&  88.03\%     	&	&87.33\%	\\
		& Thirties   &      	&  87.44\%     	&	&89.00\%	\\
		& Fourtyplus &      	&  91.82\%     	&	&91.89\%	\\
Skilled 	& 			 & 92.48\%   	&        	&92.56\%&       	\\
		& Twenties   &      	&  90.35\%    	&	&91.03\%	\\
		& Thirties   &      	&  90.22\% 	&	&92.43\%	\\
		& Fourtyplus &      	&  94.26\%     	&	&95.25\%	\\
  \hline\hline
\end{tabular}
\end{threeparttable}
\end{center}
\end{table}

The model is estimated using a \textit{flow sample} of employed and
unemployed individuals, who experienced a transition from their
original state within the period 1995-2000. We consider the first such
transition, and any subsequent transitions
are ignored. For all these individuals we can determine the beginning
of their original state, so that all durations are complete. The only
exception is constituted by a small number of individuals who disappear from the dataset
in the period 1995-2000, in which case their durations are considered
censored. We note that the period 1995-2000 was 
a period of fairly stable growth (around 2\%, with SD=.007)
and unemployment (around 8\%, with SD=.007). Focussing on
this stable period reduces the scope for biases arising from asymmetric
responses of natives and foreigners to the business cycle.

Foreigners in our sample are predominantly low
skilled: 94\% of the population of foreigners are included in our
three occupational groups, while the corresponding number for natives is approximately
86\%. The remainder occupational category is the highly skilled,
which we have excluded because of their small share in the population
of migrants (moreover, their earnings are excessively top-coded).
Table \ref{table:occup_im} considers the occupational immobility by
labour market segment. It is evident that 
occupational mobility is small, as most workers remain in
the same class. This gives further support to our segmentation
hypothesis, and such occupational immobility has also been found for
other countries (e.g. by \cite{Damas} for Portugal).

\begin{table}[htbp]
\renewcommand{\arraystretch}{.93}
\caption{Descriptives for the transition data.}\label{table:DES}
\begin{center}
\begin{tabular}{p{32pt}p{54pt}|rrr|rrr} \hline \hline
& &\multicolumn{3}{|c|}{Natives} &\multicolumn{3}{|c}{Foreigners} \\ 
Age &Transitions& Services & Unskilled & Skilled
&  Services & Unskilled & Skilled  \\ \hline
\multirow{8}{*}{25-30}& All &  8060		& 5097		& 11939		 & 1887		& 2347	  & 3023  \\                                                                                                            
&\makebox[54pt][l] {from E}				& 6088
& 3085		& 8450		 & 1438		& 1670	  & 2155 \\                                                                                          
&\makebox[54pt][r] {E$\rightarrow$ U}		& 2132		& 1764		& 4418		 & 718		& 997	  & 1225 \\                                                                                                            
&\makebox[54pt][r] {E$\rightarrow$ E}		& 3432		& 1037		& 3562		 & 373		& 351	  & 550 \\                                                                                                               
&\makebox[54pt][l] {from U}				& 1972		& 2012		& 3489		 & 449		& 677	  & 868  \\
& \makebox[54pt][r] {U$\rightarrow$ E}		& 1879		& 1932		& 3275		 & 431		& 637	  & 795 \\                                                                                                             
                
&\makebox[54pt][l] {$E_{censored}$}			& 524		& 284		& 470		 & 347		& 322	  & 380 \\                                                                                                             
&\makebox[54pt][l] {$U_{censored}$}			& 93
& 80		& 214		 & 18		& 40	  & 73 \\   
\hline
\multirow{8}{*}{30-40}& All	& 12800		& 7748
& 15381		 & 2074		& 2752	  & 3681  \\ 
&\makebox[54pt][l] {from E}				& 10723		& 5506		& 12448		 & 1637		& 2067	  & 2830 \\                                                                                          &\makebox[54pt][r] {E$\rightarrow$ U}		& 2988		& 2644		& 5284		 & 735		& 1128	   & 1451 \\                                                                                                            
&\makebox[54pt][r] {E$\rightarrow$ E}		& 6717		& 2400		& 6157		 & 453		& 477	   & 795 \\                                                                                                              
&\makebox[54pt][l] {from U}				& 2077		& 2242		& 2933		 & 437		& 685	   & 851 \\                                                                                                      
&\makebox[54pt][r] {U$\rightarrow$ E}		& 1853		& 2055		& 2601		 & 393		& 619	   & 749 \\                                                                                                             
&\makebox[54pt][l] {$E_{censored} $}		& 1018		& 462		& 1007		 & 449		& 462	   & 584 \\                                                                                                            
&\makebox[54pt][l] {$U_{censored}$}			& 224		& 187		& 332		 & 44		& 66	   & 102 \\                                                                                            \hline                                          
\multirow{8}{*}{40-55}&All	        & 16900  & 12770 & 24530   &  1494      &  2938   &5004     \\
   &\makebox[54pt][l] {from E}	        & 13912  &  9399 & 19127   &  1146      &  2090   &3726     \\ 
&\makebox[54pt][r] {E$\rightarrow$ U}	&  4538  &  4467 &  8973   &   505      &  1101   &2019     \\ 
&\makebox[54pt][r] {E$\rightarrow$ E}	&  6671  &  3206 &  6848   &   329      &   513   &1024     \\
&\makebox[54pt][l] {from U}	        &  2988  &  3371 &  5403   &   348      &   848   &1278     \\ 
&\makebox[54pt][r] {U$\rightarrow$ E}   &  1554  &  2013 &  2130   &   244      &   540   & 582     \\ 
&\makebox[54pt][l] {$E_{censored} $}	&  2703  &  1726 &  3306   &   312      &   476   & 683     \\ 
&\makebox[54pt][l] {$U_{censored}$}	&  1434  &  1358 &  3273   &   104      &   308   & 696     \\

\hline \hline                                         
\end{tabular}
\end{center}
\footnotesize{Notes: ``Censoring'' refers to a drop out
from the administrative register.}
\end{table}

Table \ref{table:DES} summarises the labour market transitions for all
nationality-age-occupation cells observed in our flow data. For both
natives and foreigners, we observe many more transitions from
employment than from unemployment. However, for
natives, the majority of transitions from employment are to another
job, whereas for the majority of foreigners the destination is
unemployment. Hence, in terms of the structural parameters, we expect
higher separation rates for foreigners, $\delta_F > \delta_N$. The
duration data for the unemployed, examined briefly in the next
subsection, suggests that foreigners exit more quickly, so that we
expect $\lambda_F > \lambda_N$ at least for this group. 

Turning to the wage data, Table \ref{table:DES3} reports for each
labour market segment the mean and standard deviation of 
wages (measured by daily gross wages in 1995
DM), as well as the average log wage gap, $\Delta \log(w) \equiv \log(w_N) - \log(w_F)$.
Natives receive substantially higher mean wages than foreigners
across all occupation groups. The segment-specific raw
  average log wage gaps in our data range from .09 to .45.  The
  overall log wage gap of .22 is in line
with reports in the literature for Germany (e.g. \cite{DustmannetalEER} report an unconditional average log wage gap of .23, \cite{Hirsch&Jahn} report a gap of .2, while \cite{Lehmer&Ludsteck}
report predicted wage gaps ranging from .08 to .44).
The three occupational groups can be
partially ordered in terms of mean wages: mean wages for the skilled 
exceed those for the unskilled for all age groups and across
nationalities. Foreign clerks and low-service workers assume an
intermediate position, but mean wages of natives in this group can
exceed those for skilled workers. 

\begin{table}[htbp]
\caption{The average wage gap in the transition data by labour market segment.}\label{table:DES3}
\begin{center}
\begin{tabular}{ll|rr|rr|rr} \hline \hline
& &\multicolumn{2}{|c|}{Services} &\multicolumn{2}{|c}{Unskilled} &\multicolumn{2}{|c}{Skilled}\\ 
Age & Wages & Native&Migrant & Native & Migrant & Native &Migrant \\ \hline

25-30 &  mean & 122.36 & 88.94 & 107.77 & 92.54 & 124.74 & 111.07 \\
& sd & 41.86 & 44.15 & 37.68 & 36.09 & 29.94 & 35.21 \\
& $\Delta \log(w)$ & \multicolumn{2}{|c}{.32} &
\multicolumn{2}{|c}{.15} & \multicolumn{2}{|c}{.11} \\ \hline

30-40 & mean & 156.35 & 99.38 & 120.94 & 97.99 & 135.79 & 116.65 \\
& sd & 51.22 & 55.02 & 38.24 & 36.61 &32.04 & 36.22 \\
& $\Delta \log(w)$ & \multicolumn{2}{|c}{.45} &
\multicolumn{2}{|c}{.21} & \multicolumn{2}{|c}{.15} \\ \hline

40-55 & mean & 158.17 & 112.74 & 125.05 & 107.49 & 138.29 & 126.20 \\
& sd & 48.09 & 56.81 & 36.71 & 36.89 & 33.29 & 33.50 \\
& $\Delta \log(w)$ & \multicolumn{2}{|c}{.33} & \multicolumn{2}{|c}{.15} & \multicolumn{2}{|c}{.09} \\
\hline \hline                                         
\end{tabular}
\end{center}
\footnotesize{Notes: $\Delta \log(w) \equiv \log(w_N) -
  \log(w_F)$. The overall log wage gap is .22. Wage dating: for transitions from employment (E$\rightarrow$ \{U,E\}), these are the last earned wages in this state,
for transition out of unemployment (U$\rightarrow$ E) these are the
first wages earned in the new job.}
\end{table}

\begin{figure}[h]
\centering
\caption{Estimates of the density of accepted wages by labour market
  segments.}
\includegraphics[scale=0.4] {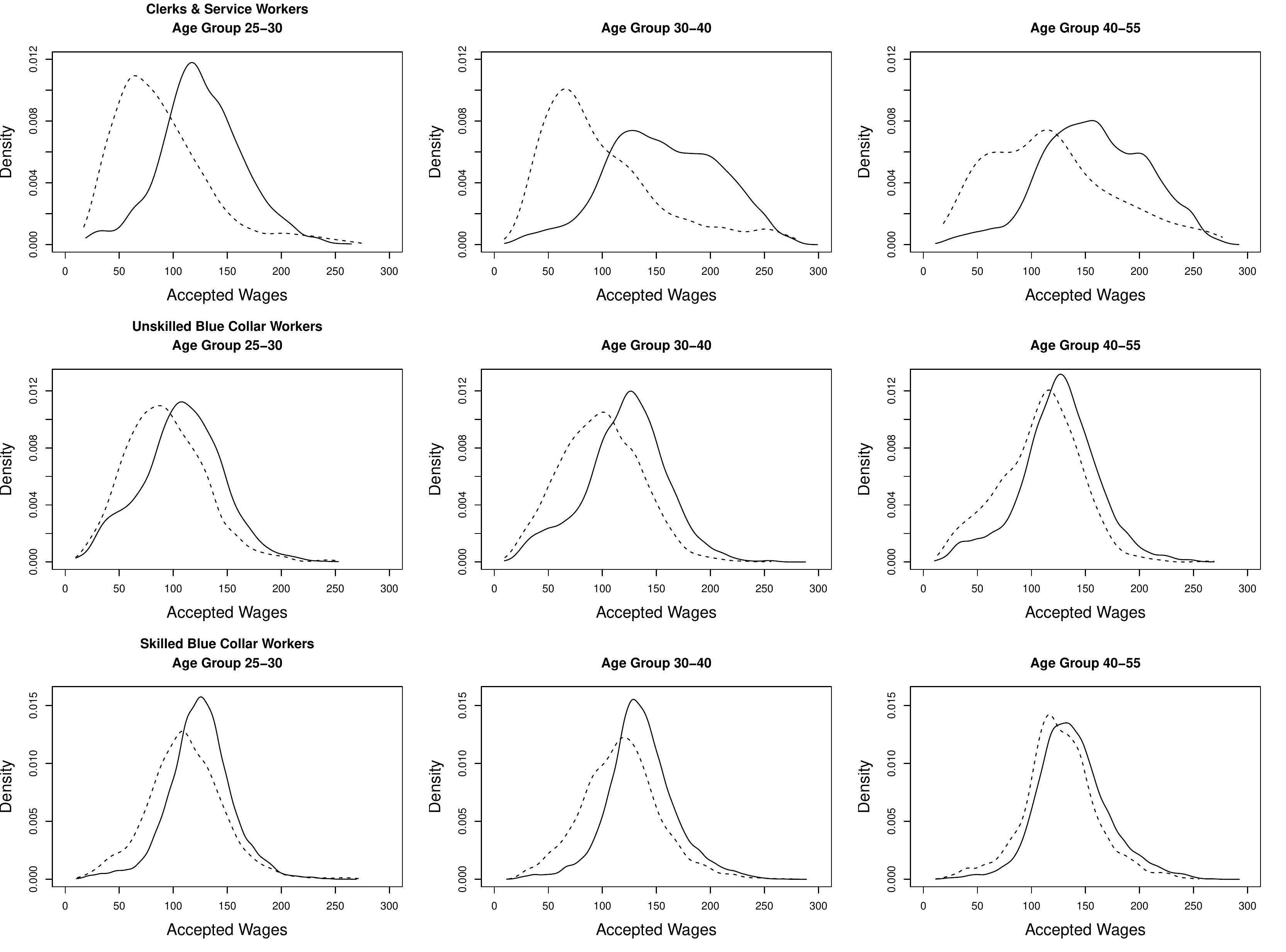}\label{fig:wage_density}\\
{\footnotesize Notes: Natives (solid lines) v. foreigners (dashed lines).}
\end{figure}

Rather than only restricting
attention to the mean wage, Figure \ref{fig:wage_density} depicts the
kernel estimates of the realised wage densities (the solid lines refer
to natives). The most pronounced distributional difference exist for
the semi-skilled workers (clerks and service workers), and the differences persist across age
groups. By contrast, for all other occupations, the differences
decrease in age. The density estimates also exhibit ``blips'' in the far
left tails of the wage densities. This bimodality leads to problems in
the estimation of the model, manifesting themselves by the occurrence
of spikes in the estimated productivity density. We overcome this
issue by truncating the wage distributions
at the 5\% percentile, which is a common cut-off in the literature
(see e.g. \cite{Bowlus1997} or \cite{FlabbiIER}). The estimation of the
reservation wage distribution is, of coure, likely to be sensitive to
the choice of the cut-off point. We therefore explore the robustness
of our parameter estimates below in Section \ref{sec:robustness}, and
find that the frictional parameters are fairly stable, while $\mu$ increases
usually somewhat as the truncation increases from 3\% to 7\%.

\subsubsection{Reduced Form Estimates: The Importance of Unobservable
  Heterogeneity}

Before embarking on the estimation of the model, we first explore
descriptively whether there is scope for unobserved heterogeneity to
play a role in explaining unemployment durations. To this end, we
estimate standard reduced-form proportional hazard (PH) and mixed
proportional hazard (MPH) models for the
unemployed, controlling incrementally for duration dependence and 
unobserved heterogeneity. Since the conditional unemployment durations
in the structural model are exponential with parameter $\lambda
\bar{F}(b)$ and the marginal durations are a mixture of such
exponentials, we first estimate an exponential PH model, and then
allow for duration dependence by estimating a Weibull
specification. As the latter confounds dynamic sorting driven by
unobservable heterogeneity and genuine duration dependence 
(see e.g. \cite{VanDenBerg2001}), we then estimate MPH models using the
common gamma frailty (assumed to be independent of the
covariates). Note, however, that these reduced-form
parameters do not identify the parameters of the structural model as
the former are complicated functions of the latter. In
all models we condition on interactions between age and occupational
groups in order to mirror our subsequent structural analysis of the
corresponding labour market segments.

Table \ref{table:Hazards} reports the results. Across all models the
migrant dummy is positive throughout, so that their job offer
arrival rates exceed those of natives. The Weibull PH model suggests
the presence of duration dependence, but the MPH reveals this to be
caused by dynamic sorting: once unobservable heterogeneity is
controlled for, the Weibull parameter does not differ statistically
from 1. Hence Weibull and exponential MPH models yield similar
coefficient estimates.
This inferred absence of duration dependence is consistent
with the structural model, as it cannot generate genuine duration
dependence but does yield dynamic sorting through unobserved heterogeneity in
reservation wages.
 
\begin{table}[htbp]
\begin{center}
\begin{threeparttable}
\small
\caption{Reduced-form unemployment duration models}\label{table:Hazards}
\def\sym#1{\ifmmode^{#1}\else\(^{#1}\)\fi}
\begin{tabular}{l*{5}{c}} \hline\hline
            &&\multicolumn{1}{c}{(1)}&\multicolumn{1}{c}{(2)}&\multicolumn{1}{c}{(3)}&\multicolumn{1}{c}{(4)}\\
            &&\multicolumn{1}{c}{Exponential}&\multicolumn{1}{c}{Weibull}&\multicolumn{1}{c}{Weibull$^{\S}$}
            &\multicolumn{1}{c}{Exponential$^{\S}$} \\  \hline
\multicolumn{2}{l}{Migrant}                    & .087$^{***}$ & .069$^{***}$ & .049$^{*}$ & 0.046$^{*}$ \\
                      &    & (.020) & (.020) & (.027) & (.027) \\
\multicolumn{2}{l}{Clerks $\times$ Twenties}     & 1.368$^{***}$ & 1.212$^{***}$ & 1.409$^{***}$ &  1.431$^{***}$  \\ 
                       &   & (.035) & (.036) & (.047)  & (.047) \\
\multicolumn{2}{l}{Clerks $\times$ Thirties}    & 1.059$^{***}$ & .984$^{***}$ & 1.197$^{***}$ &   1.217$^{***}$  \\ 
                       &     & (.034) & (.035) & (.047) &  (.046)\\
\multicolumn{2}{l}{Clerks $\times$ Fourtyplus}  &.037 & .011 & -0.005 &    -0.006 \\
                        &     & (.037) & (.037) & (.045) &  (.046) \\    
\multicolumn{2}{l}{Skilled $\times$ Twenties}  & 1.500$^{***}$ & 1.327$^{***}$ & 1.602$^{***}$ &   1.631$^{***}$ \\
                        &    & (.031) & (.031) & (.043) &  (.041) \\
\multicolumn{2}{l}{Skilled $\times$ Thirties}   & .914$^{***}$ & .867$^{***}$ &  1.155$^{***}$ &  1.182$^{***}$   \\
                        &    & (.032) & (.033) & (.045) &  (.044) \\
\multicolumn{2}{l}{Skilled$\times$ Fourtyplus}   & -0.429$^{***}$ & -0.451$^{***}$& -0.531$^{***}$ &  -0.539$^{***}$ \\
                        &      & (.034) & (.034) & (.041) & (.042) \\
\multicolumn{2}{l}{Unskilled$\times$ Twenties}  & 1.297$^{***}$ & 1.153$^{***}$ & 1.392$^{***}$  &  1.417$^{***}$   \\
                        &       & (.035) & (.035) & (.047)&  (.046) \\
\multicolumn{2}{l}{Unskilled $\times$ Thirties}   &.864$^{***}$ & .828$^{***}$ & 1.062$^{***}$ &  1.083$^{***}$  \\
                        &     & (.035) & (.034) & (.046) & (.046) \\ \hline
duration &\multirow{2}{*}{$ln(\alpha)$}    & & -.222$^{***}$ &  -.023$^{*}$ & \\
dependence &                   & & (.006) & (.012)  & \\ \hline
unobserved &\multirow{2}{*}{$\theta$}        & & &.702$^{***}$  &  .770$^{***}$ \\
heterogeneity &                & & &( .042)  &  (.024)\\
\hline\hline
\end{tabular}
\begin{tablenotes}
\item Notes. Standard errors in parentheses, $^* (p<0.1)$, $^{***} (p<0.001)$.
 Reference groups: Unskilled $\times$ Fourtyplus. $^{\S}$Frailty is Gamma distributed.\\
\end{tablenotes}
\end{threeparttable}
\end{center}
\end{table}

\section{Estimation Results}\label{results_section}

We proceed to estimate the structural parameters of the model,
i.e. the job offer arrival rate, $\lambda$, the match destruction
rate, $\delta$, and the parameters of the distribution of workers'
reservation values, $(\mu, \sigma)$, as well as the density of firms' productivity
in each segment. Each occupation group is considered in
turn, and we segment for each occupation the labour market further by age
and nationality. The average migrant effects and the wage decompositions are
then quantified in detail in Section \ref{sec:experiments} below.

\begin{table}[htbp]
\begin{center}
\begin{threeparttable}
\caption{Structural parameter estimates: Unskilled blue collar workers} \label{table:UBCW_A_1}
\begin{tabular}{c|c|cccc|c}
\hline \hline
Age & Nation. &  $\mu$& $\sigma$& $\lambda$&$\delta$ &    $k=\lambda/\delta $        \\\hline

\multirow{4}{*}{\textbf{25-30}} &N
                                       &  53.76 & 11.10 & .0666 & .0257               &  2.59     \\
							        & &[51.74-56.02]&[9.67-14.06]&[.0487-.0891]&[.0241-.0268] &       \\ 		
                             &F  & 50.15 & 17.47 & .1705 & .0339                       &  5.03     \\ 
 		            	& &[46.91-52.04]&[14.86-20.34]&[.1447-.1932]&[.0307-.0358] &       \\ 		
				\hline
\multirow{4}{*}{\textbf{30-40}}
                             &N  & 50.97 & 8.76 & .0416 & .0098                               & 4.24        \\ 
                                  & &[49.06-53.77]&[6.95-11.10]&[.0356-.0583]&[.0092-.0106] &       \\ 		 
                     &F    & 49.35 & 15.86 & .1071 & .0167                             & 6.41   \\ 
				& &[46.33-50.78]&[13.14-20.2]&[.0762-.1261]&[.0162-.0178]   &   \\ 
				\hline
\multirow{4}{*}{\textbf{40-55}}
				&N& 54.05 & 10.10 & .0355 & .0051                           & 6.96 \\ 
			& &[51.81-55.98]&[8.56-11.87]&[.0281-.0412]&[.0048-.0056] &       \\ 		
				&F & 50.44 & 8.12 & .0353 & .0072                            &4.90\\ 
				& &[47.62-52.88]&[6.40-11.92]&[.0221-.0501]&[.0067-.0075] &  \\
										
\hline\hline
\end{tabular}
\begin{tablenotes}[normal,flushleft]
\item \footnotesize Notes: In brackets: the 2.5\% and 97.5\% percentiles of the bootstrap distribution.
\end{tablenotes}
\end{threeparttable}
\end{center}
\end{table}

\subsection{Unskilled Blue Collar Workers}

Table \ref{table:UBCW_A_1} reports the results. Across all three age
groups, the labour turnover parameters of migrants exceed those of
natives, $\hat{\delta}_F > \hat{\delta}_N$ and $\hat{\lambda}_F >
\hat{\lambda}_N$. Migrants experience job separations more often, but
this is partially compensated by them also finding new jobs more
quickly. All job turnover parameters fall in age.
Across age groups and nationality, transitions into new
jobs happen more quickly than transitions into unemployment, $\hat{\lambda}
> \hat{\delta}$. Foreigners have slightly lower mean reservation wages,
$\hat{\mu}_F < \hat{\mu}_N$, but confidence intervals overlap. The
estimates are fairly stable across age. 
The estimates for the reservation wage distribution for both groups
imply that not all new job offers are accepted: there are some workers with high reservation wages who
would and do turn down new job offers with insufficiently high wages.

\begin{figure}[htbp]
\centering
\caption{Unskilled blue collar workers aged 25-30.}
\includegraphics[scale=0.35] {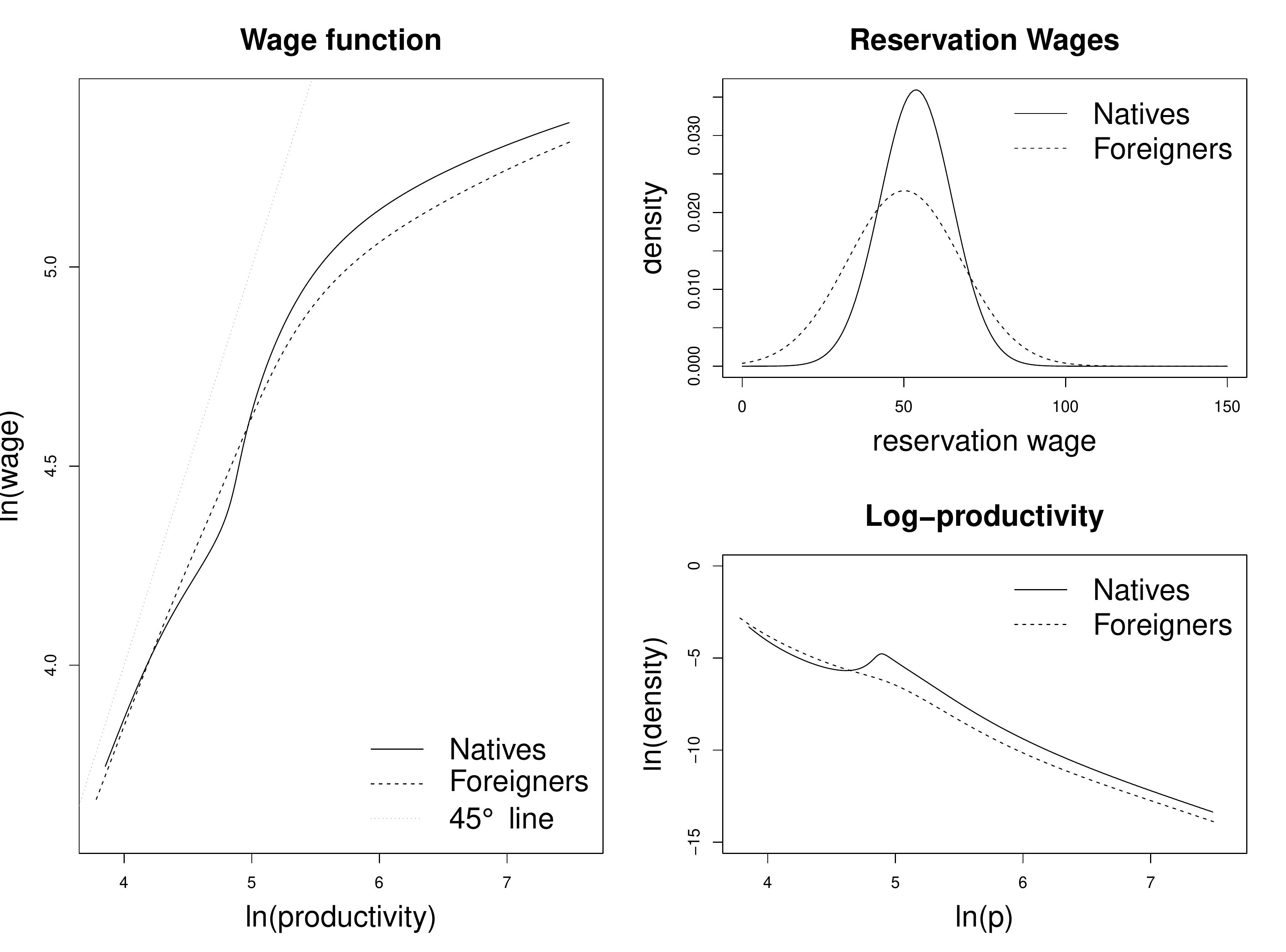}
\label{fig:young89}
\end{figure}

In Figure~\ref{fig:young89} we consider some implications of the
estimated model for the young. Panel A plots the wage offer functions, 
panel B the reservation wage density, whilst panel C plots the estimated
productivity densities\footnote{These are obtained as follows. Given the
parameter estimates and kernel estimate of the realised wage density,
the unemployment rate $u$ is estimated using equation (\ref{SS_A2}),
and the wage offer distribution $F$ follows from 
(\ref{SS_A1}); the productivity distribution is then estimable from
equation (\ref{Prod_A}).}. It is evident that the productivity
densities for both groups are well approximated by a Pareto
density. The slopes for sufficiently high productivities are very
similar. Turning to wage offers (panel A), for low productivities
foreigners do not do worse than natives, while for log productivities
above 5 natives receive better wage offers. Overall, the figure
suggests a positive but small migrant effect, and this is confirmed by
our quantifications reported in Section \ref{sec:experiments}.

\subsection{Clerks and Low-Service Workers}

\begin{table}[htbp]
\begin{center}
\begin{threeparttable}
\caption{Structural parameter estimates: Clerks \& service workers} \label{table:UBCW_A_2}
\begin{tabular}{c|c|cccc|c}
\hline \hline
Age & Nation. &  $\mu$& $\sigma$& $\lambda$&$\delta$ &   $k=\lambda/\delta $        \\\hline

\multirow{4}{*}{\textbf{25-30}} 
& N & 65.60 & 14.39 & .0984 & .0194				&5.07	 \\ 
&  &[61.92-66.53]&[11.61-15.8]&[.0697-.0836]&[.0189-.0199] &		 \\
 & F   & 36.09 & 13.65 & .0701 & .0272 		    &	2.58	\\ 
& &[30.88-41.69]&[8.6-17.17]&[.0624-.0886]&[.0259-.0284] & 		 \\ 
\hline
\multirow{4}{*}{\textbf{30-40}} 
&N  & 72.66 & 9.42 &.0423 &.0073			&5.79		 \\ 
& &[68.41-75.12]&[7.54-10.39]&[.0355-.0530]&[.0071-.0076] &       \\ 	
& F & 43.27 & 7.40 & .0593 & .0157			&3.77		 \\
& &[40.88-45.62]&[6.41-9.57]&[.0478-.0703]&[.0151-.0162] &       \\ 	
\hline
\multirow{4}{*}{\textbf{40-55}} 
&N   & 73.07 & 7.92 & .0698 & .0035			&19.94		 \\ 
& &[70.51-75.12]&[7.07-9.16]&[.0603-.0841]&[.0031-.0037] &       \\ 		
&F  & 49.04 & 6.86 &.0759 & .0077			&9.86		 \\ 
& &[46.38-51.94]&[5.22-8.41]&[.0565-.0911]&[.0072-.0081] &       \\ 		

\hline\hline
\end{tabular}
\begin{tablenotes}[normal,flushleft]
\item \footnotesize Notes: As for Table \ref{table:UBCW_A_1}.
\end{tablenotes}
\end{threeparttable}
\end{center}
\end{table}

\begin{figure}[htbp]
\centering
\caption{Clerks and service workers aged 25-30.}
\includegraphics[scale=0.35] {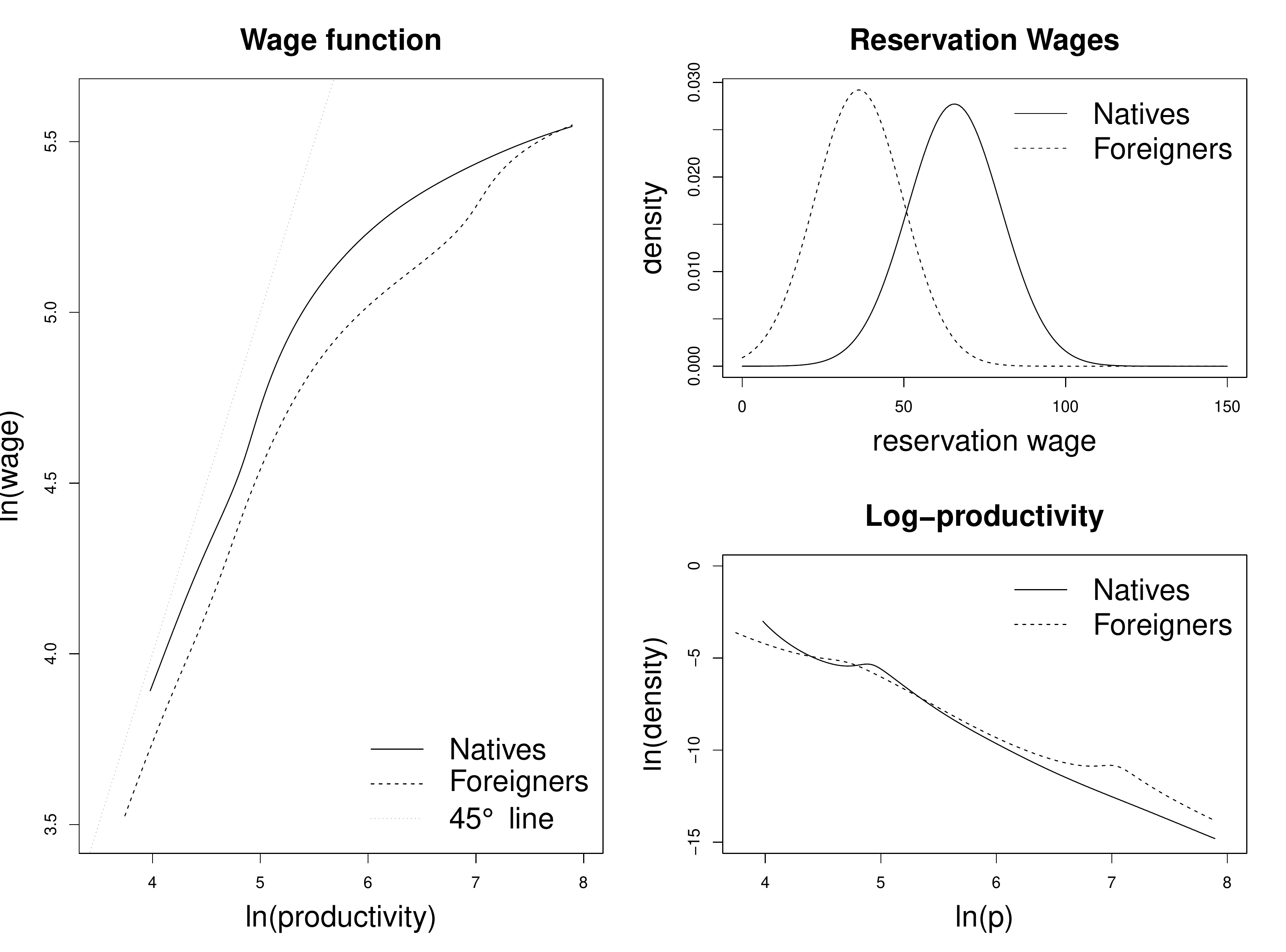}
\label{fig:young345}
\end{figure}

Table \ref{table:UBCW_A_2} reports the results for this occupational
group, for which we observed in Table \ref{table:DES3} the largest
average wage gap. As before, job separation rates for foreigners
exceed those of natives, decline in age, and are smaller than job
offer arrival rates. Except for the young, the transition rates of
foreigners exceed those of natives. But unlike the case of the
unskilled, differences in mean reservation wages are substantial:
foreigners are substantially less demanding, on average, than
natives. These means increase in age.
Figure \ref{fig:young345} panel C suggests that productivities are again well approximated by a Pareto
form, and panel A suggests that the maximal migrant effect is
substantial.

\subsection{Skilled Blue-Collar Workers}

\begin{table}[htbp]
\begin{center}
\begin{threeparttable}
\caption{Structural parameter estimates: Skilled blue collar workers} \label{table:UBCW_A_3}
\begin{tabular}{c|c|cccc|c}
\hline \hline
Age & Nation. &  $\mu$& $\sigma$& $\lambda$&$\delta$ &  $k=\lambda/\delta $        \\\hline

\multirow{4}{*}{\textbf{25-30}}&N    & 81.15 & 4.52 & .0801 & .0158                        &5.07      \\ 
&                                    & [77.64-83.97]&[3.81-6.59]&[.0684-.0911]&[.0121-.0179] &             \\
 & F                               & 66.38 & 14.05 & .1067 & .0225                            &4.74            \\ 
&                                    & [62.88-69.04]&[11.88-17.32]&[0.955-0.1182]&[.0170-.0268] &           \\   
 \hline

\multirow{4}{*}{\textbf{30-40}}& N & 76.68 & 8.85 & .0698 & .0068                       &10.26                \\ 
                                 & &[73.82-77.90]&[7.69-9.71]&[.0621-.0760]&[.0063-.0071] &       \\
& F                                &  69.30 & 8.06 & .0866 & .0124                   & 6.98\\  
                                  & &[65.57-72.05]&[7.34-9.16]&[.0681-.0946]&[.0119-.0127] &       \\ 
\hline 
\multirow{4}{*}{\textbf{40-55}}& N  & 79.71 & 6.44 & .0408 & .0035                      &11.66\\ 
                                 & &[77.18-80.94]&[5.62-7.01]&[.0343-.0478]&[.0033-.0036] &       \\ 
                                & F  & 75.05 & 7.33 & .0449 & .0049                      &9.16\\ 
                                 & &[71.48-78.24]&[6.51-8.70]&[.0325-.0512]&[.0045-.0051] &       \\ 	
\hline\hline
\end{tabular}
\begin{tablenotes}[normal,flushleft]
\item \footnotesize Notes: As for Table \ref{table:UBCW_A_1}.
\end{tablenotes}
\end{threeparttable}
\end{center}
\end{table}

\begin{figure}[htbp]
\centering
\caption{Skilled blue collar workers aged 25-30.}
\includegraphics[scale=0.35] {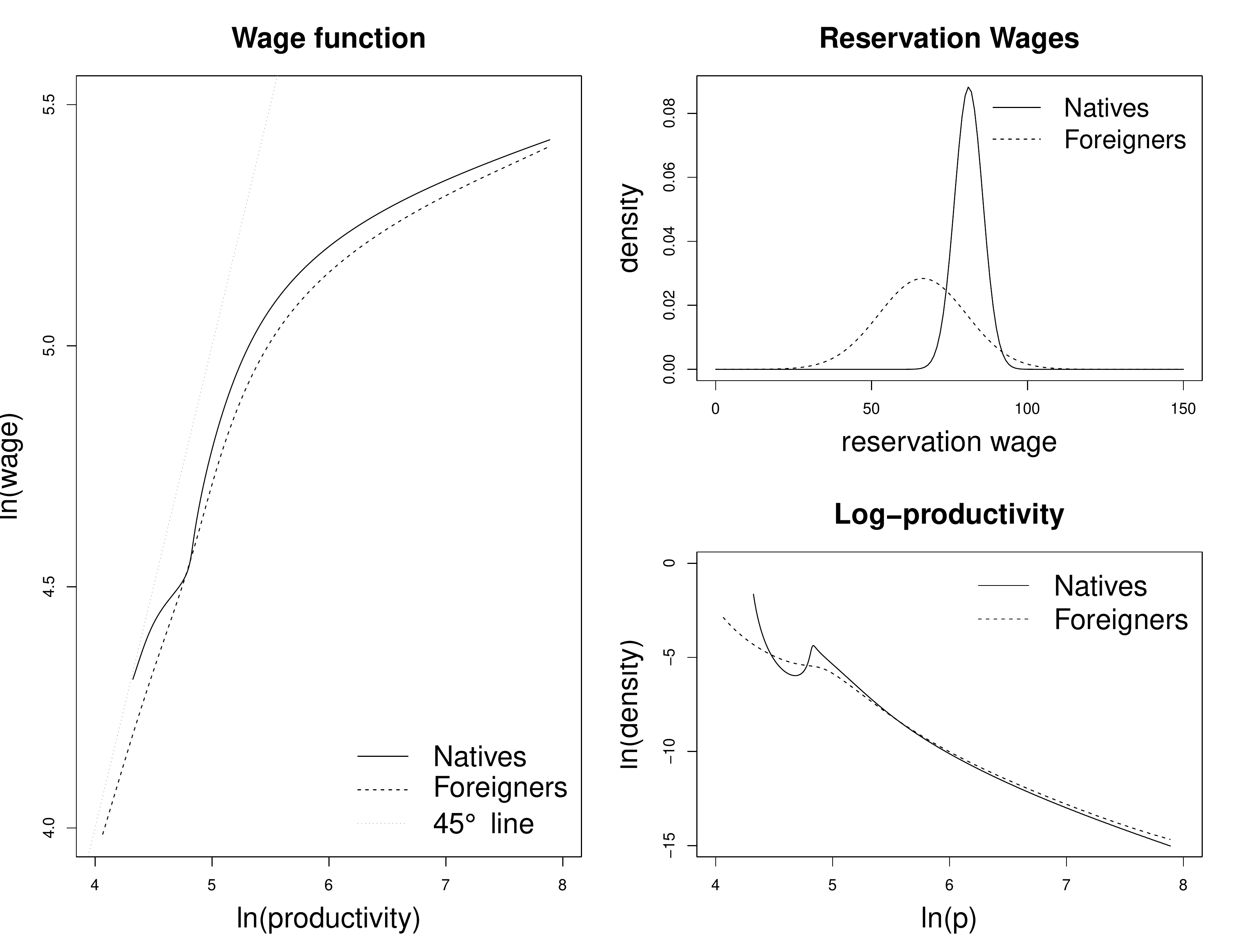}
\label{fig:young7}
\end{figure}

For the skilled blue-collar workers, the by now familiar pattern
emerges too, as is evident from Table \ref{table:UBCW_A_3}: both turnover
parameters are higher for migrants, and decline in age. As regards mean
reservation wages, foreigners are less demanding than natives, but the
gap is not as wide as for clerks and service workers, and it declines
in age. Focussing on the young in Figure
\ref{fig:young7}, productivities are Pareto like. The
migrant effect, captured in Panel A, is modest.

\subsection{General Discussion}
 
Comparing the results across occupations, we observe similar
patterns. Migrants experience job separations more often than natives
but also typically find jobs more quickly, and job turnover parameters
tend to decline
in age. These findings are in line with differences in employment
protection observed in \cite{Sa2011}, who
reports that migrants in Germany are much more likely than natives
to work on temporary contracts. The findings are also consistent with the other dimensions
of segregation extensively documented in \cite{Glitz2012}. 
Across all segments and nationality, transitions into new
jobs happen more quickly than transitions into unemployment.
Overall, search frictions, as measured by
  $(\lambda/\delta)^{-1}$, are of the same order of magnitude across all occupational
  groups, decrease in age (except for unskilled foreigners
  in which case they are stable), and are larger for foreigners than
  for natives for the skilled and clerks and service workers. Thus,
  the higher job offer arrival rate for foreigners cannot compensate
  for their higher job separation rates.
As regards the reservation wage distribution, 
across all segments there are some workers with high reservation
wages who turn down new job offers when wage offers are too
low.\footnote{These results differ from estimates for Netherlands
(\cite{VanDenBerg&Ridder}) and France (\cite{Bontempsetal1999})
since both countries have a binding legal minimum wage. Similar to
these studies, however,  we observe that job separation parameter
$\delta$ is approximately one order of magnitude
smaller than the estimated job offer arrival rate $\lambda$.
Our results are comparable to those reported by
\cite{BartolucciGender} for Germany obtained from a different empirical
search model applied to a different market segmentation. For low
qualified male workers in the manufacturing sector he reports a job
separation rate of .03 and a job offer arrival rate of .3.  Our
estimates of $\delta$ range from .004 to .03, and the job offer
arrival rate ranges from .04 to .17. Our result regarding the mean reservation wage, the lack of a statistically significant difference between natives and migrants for most groups, is also consistent with external evidence reported in \citet[Table~2]{Bergemannetal} who use the IZA evaluation sample of individuals who entered unemployment in late 2007 and early 2008: the means of the self-reported reservation wages are not statistically significantly different between natives and migrants, and their levels are comparable to ours.} Migrant
workers are on average less demanding than natives.
Firm productivities are well approximated by Pareto forms.
However, migrants
receive wage offers that are lower than for natives who have the same
productivity. This migrant effect is the largest for clerks and service
workers, and small for unskilled workers. The drivers of the migrant
effect are the subject of Section \ref{sec:experiments}.

\subsection{Robustness Checks}\label{sec:robustness}

\subsubsection{Return Migration}

A concern for our estimates in the migrant segments might
  be the effect of return migration, when such returnees leave Germany
  out of employment. In order to investigate the sensitivity of our
  estimates to this issue, we consider restricted samples of migrants
  who should, in principle, be available for work after their
  employment transition, by requiring foreigners to be observable
  in the data 6 months after their transition. This restriction leads
  to a net dropout of foreigners (relative to that of natives) across segments between 7.7\%
  and 15.3\%.\footnote{These rates are consistent with
      \citet[Table~1]{Gundel&Peters} who, using GSOEP data for the
      period 1984-2005, suggest that among male immigrants aged less than
      60 the return rate is 10\%} Table
  \ref{table:Return_mig} juxtaposes the estimates for these restricted
  samples  (labelled noRetMig) to our unrestricted estimates. We find
  that most parameter estimates remain relatively stable (the occasional fall in $\hat{\sigma}$ reflects the extent to which the sample restriction increases the homogeneity of the group; the more homogeneous the sample, the smaller the estimated reservation wage dispersion).

\begin{table}[htbp]
\begin{center}
\begin{threeparttable}
\caption{Sensitivity analysis - the effect of excluding return migrants} \label{table:Return_mig}
\begin{tabular}{c|c|c|cccc}
\hline \hline
 Occupation &  Age & Group &  $\mu$& $\sigma$& $\lambda$&$\delta$ \\\hline

 \multirow{6}{*}{Unskilled} &\multirow{2}{*}{25-30}  & full     &50.15 & 17.47 & 0.1705 & 0.0339 \\ 
                      &                             & noRetMig & 49.88 & 10.14 & 0.1207 & 0.0433 \\ 
\cline{2-7}              
                      &\multirow{2}{*}{30-40}       & full     &49.35 & 15.86 & 0.1071 & 0.0167 \\
                      &                             & noRetMig & 50.65 & 8.53 & 0.0588 & 0.0219 \\
\cline{2-7}
                      & \multirow{2}{*}{40-55}      & full     & 50.44 & 8.12 & 0.0353 & 0.0072 \\ 
                      &                             & noRetMig & 48.53 & 3.28 & 0.0495 & 0.0087 \\ 

  \hline\hline   
\multirow{6}{*}{Skilled} &\multirow{2}{*}{25-30}     & full     &66.38 & 14.05 & 0.1067 & 0.0225 \\ 
                     &                               & noRetMig & 65.00 & 9.05 & 0.0871 & 0.0281 \\ 
\cline{2-7}
                     &\multirow{2}{*}{30-40}     & full     &69.30 & 8.06 & 0.0866 & 0.0124 \\ 
                     &                               & noRetMig & 68.72 & 5.86 & 0.0660 & 0.0142 \\ 
\cline{2-7}
                      & \multirow{2}{*}{40-55} & full     & 75.05 & 7.33 & 0.0449 & 0.0049 \\ 
                      &                        & noRetMig &  62.74 & 7.44 &  0.0379 & 0.0058 \\
  \hline\hline   
                                     &\multirow{2}{*}{25-30} & full     &36.09 & 13.65 & 0.0701 & 0.0272 \\
                                     &                       & noRetMig &37.84 & 11.37 & 0.0749 & 0.0344 \\
\cline{2-7}
           Clerks                    &\multirow{2}{*}{30-40} & full     &43.27 & 7.40 & 0.0593 & 0.0157 \\ 
           \& Services               &                       & noRetMig &45.11 & 9.09 & 0.0689 & 0.0201 \\ 
\cline{2-7}                                                                           
                                     & \multirow{2}{*}{40-55} & full     &49.04 & 6.86 & 0.0759 & 0.0077 \\ 
                                     &                        & noRetMig &48.23 & 5.39 & 0.1036 & 0.0098 \\  
   \hline\hline   
\end{tabular}
\end{threeparttable}
\end{center}
\end{table}

\subsubsection{The effect of truncating the wage distribution}
\begin{table}[htbp]
\renewcommand\tabcolsep{4pt}
\begin{center}
\begin{threeparttable}
\caption{Sensitivity analysis -  the effects of truncation} \label{table:Trunc}
\begin{tabular}{c|c|c|cccc|cccc}
\hline \hline
 Occupation &  Age & Trunc. & \multicolumn{4}{c|}{Foreigners} &
 \multicolumn{4}{c}{Natives} \\ 
& & &  $\mu$& $\sigma$& $\lambda$&$\delta$ &
 $\mu$& $\sigma$& $\lambda$&$\delta$ \\\hline
\multirow{9}{*}{Unskilled} & \multirow{3}{*}{25-30}       & 3\% &  45.31 & 19.68 & .1698 & .0344  & 42.77 & 15.85 & .0603 & .0256\\ 
                           &                             & 5\% &  50.15 & 17.47 & .1705 & .0339 &  53.76 & 11.10 & .0666 & .0257\\ 
                           &                             & 7\% &  54.47 & 15.18 & .1593 & .0333 &  56.72 & 9.53 & .0783 & .0254\\ 
\cline{2-11}                                                        
                           &\multirow{3}{*}{30-40}       & 3\% &   47.89 & 16.51 & .1215 & .0167 &  38.62 & 11.64 & .0321 & .0099\\
                           &                             & 5\% &   49.35 & 15.86 & .1071 & .0167 &  50.97 & 8.76 & .0416 & .0098\\
                           &                             & 7\% &   55.62 & 12.56 & .1000 & .0162 &  57.53 & 9.10 & .0306 & .0095\\
\cline{2-11}                                                        
                           & \multirow{3}{*}{40-55}      & 3\% &   40.18 & 3.86 & .0435 & .0074 &  38.62 & 11.64 & .0321 & .0099\\ 
                           &                              & 5\% &   50.44 & 8.12 & .0353 & .0072 &  54.05 & 10.10 & .0355 & .0051\\ 
						   &
                                                   & 7\% &   52.83 &
                                                   7.64 & .0298 &
                                                   .0071 &   53.87 &
                                                   14.73 & .0276 &
                                                   .0049\\ \hline \hline

\multirow{9}{*}{Skilled} &\multirow{3}{*}{25-30}       & 3\% & 57.41 & 18.69 & .0915 & .0229 &   72.78 & 9.66 & .0611 & .0162\\ 
                         &                             & 5\% & 66.38 & 14.05 & .1067 & .0225 & 81.15 & 4.52 & .0801 & .0158\\ 
                         &                             & 7\% & 71.79 & 10.38 & .1061 & .0219 & 82.65 & 3.72 & .0729 & .0154\\ 
\cline{2-11}                                                               
                         &\multirow{3}{*}{30-40}       & 3\% & 63.53 & 12.14 & .0695 & .0127 & 73.51 & 9.44 & .0798 & .0069\\ 
                         &                             & 5\% & 69.30 & 8.06 & .0866 & .0124 & 76.68 & 8.85 & .0698 & .0068\\ 
                         &                             & 7\% & 72.36 & 6.56 & .0579 & .0121 & 87.08 & 4.58 & .0612 & .0064\\ 
\cline{2-11}                                                               
                         & \multirow{3}{*}{40-55}      & 3\% &  67.90 & 7.57 & .0557 & .0045 &  69.17 & 8.12 & .0407 & .0036\\ 
                         &                              & 5\% & 75.05 & 7.33 & .0449 & .0049  & 79.71 & 6.44 & .0408 & .0035\\ 
                         &            		          & 7\% & 77.32 & 5.81 & .0392 & .0045 & 83.31 & 5.40 & .0583 & .0035\\ 
  \hline\hline   
                              &\multirow{3}{*}{25-30}       & 3\% & 35.43 & 14.04 & .0628 & .0269 & 58.30 & 18.68 & .1113 & .0198\\
                               &                             & 5\% & 36.09 & 13.65 & .0701 & .0272 & 65.60 & 14.39 & .0984 & .0194\\        
                               &                             & 7\% & 35.44 & 14.04 & .0628 & .0269 & 68.11 & 13.14 & .0953 & .0191\\
 \cline{2-11}                                                                       
                               &\multirow{3}{*}{30-40}       & 3\% & 41.98 & 7.83 & .0608 & .0156 & 62.80 & 15.13 & .0555 & .0075\\ 
Clerks                         &                             & 5\% & 43.27 & 7.40 & .0593 & .0157 & 72.66 & 9.42 & .0423 & .0073\\ %
\& Services                    &                             & 7\% &
47.34 & 5.79 & .0464 & .0154 & 78.16 & 5.92 & .0490 & .0072 \\
\cline{2-11}                                                                         
                               & \multirow{3}{*}{40-55}      & 3\% & 47.89 & 8.88 & .0534 & .0078 & 66.87 & 10.03 & .0452 & .0035\\  
                               &                              & 5\% &
                               49.04 & 6.86 & .0759 & .0077 & 73.07 & 7.92 & .0698 & .0035 \\ 
                               &            		         & 7\% & 48.59 & 5.09 & .0635 & .0076 & 78.09 & 6.47 & .0709 & .0034\\ 
   \hline\hline   
\end{tabular}
\end{threeparttable}
\end{center}
\end{table}
Our samples have been truncated at 5\% at the left tail
  of the wage distribution, a common cut-off in the literature. Here,
  we examine the sensitivity of our estimates to varying the cut-off
  from 3\% to 7\%. Table \ref{table:Trunc} reports the results. Across
  all segments, the frictional parameters $\delta$ and $\lambda$ are
  very stable. An increase in the truncation is expected to lead to an
  increase in the estimated mean reservation wage. This increase, however, turns
  out to be typically very modest. We conclude that our estimates are
  robust.

\subsubsection{Ethnic German Immigrants} \label{ethnic_Germans}
\begin{table}[htbp]
\begin{center}
\begin{threeparttable}
\caption{Native workers: full and restricted sample results} \label{table:Robust}
\begin{tabular}{c|c|c|cccc}
\hline \hline
 Occupation &  Age & Group &  $\mu$& $\sigma$& $\lambda$&$\delta$ \\\hline

 \multirow{4}{*}{Unskilled} &\multirow{2}{*}{30-40} & all & 50.97 & 8.76 & .0416 & .0098 \\ 
                      &                       & pre '88&48.46 & 5.22 & .0362 & .0090 \\ \cline{2-7}

                      & \multirow{2}{*}{40-55} & all &54.05 & 10.10 & .0355 & .0051 \\ 
 
                      &                        & pre '88& 51.95 & 6.66 & .0203 & .0046 \\ 

  \hline\hline   
\multirow{4}{*}{Skilled} &\multirow{2}{*}{30-40} & all  &76.68 & 8.85 & .0698 & .0068 \\ 
                     &                       & pre '88&88.53 & 2.78 & .0700 & .0060 \\ 
\cline{2-7}
                      & \multirow{2}{*}{40-55} & all &79.71 & 6.44 & .0408 & .0035 \\ 
                      &                        & pre '88&81.34 & 7.51 & .0407 & .0032 \\ 
  \hline\hline   
\multirow{2}{*}{Clerks} &\multirow{2}{*}{30-40} & all  &72.66 & 9.42 & .0423 & .0073 \\   
 \multirow{2}{*}{ \& Services}                   &                       &  pre '88& 71.11 & 9.12 & .0496 & .0067 \\ 
\cline{2-7}
                      & \multirow{2}{*}{40-55} & all &73.07 & 7.92 & .0698 & .0035 \\  
                      &                        & pre '88& 72.66 & 8.93 & .0413 & .0033 \\
 
   \hline\hline   
\end{tabular}
\begin{tablenotes}[normal,flushleft]
\item \footnotesize Notes: ``all'' refers to the full sample of native
  workers, ``pre '88'' to the sample of natives observed before 1988.
\end{tablenotes}
\end{threeparttable}
\end{center}
\end{table}

The inflow of foreign-born ethnic Germans in the late 1980's and early
1990's changed the composition of the group of natives. While
qualifying for a German passport by descent, many did not speak German
and were more similar to the group of foreign nationals considered
above. However, these ethnic German immigrants are not directly identifiable in
our data and thus latent in the group of natives. This arguable
misclassification could lead to biases in our estimates for native
workers. To check the robustness of our results to such changes in the
population of German citizens, we estimate the model using
the subsample of native workers that are also present in the data set before
1988 (labelled pre'88), the year before the inflow of ethnic Germans occurred. Table
\ref{table:Robust} reports our estimates, and for ease of comparison,
juxtaposes these to our earlier results for the unrestricted sample
(labelled all). The young age group is excluded from this exercise
since many in this group would be too young to be employed pre 1988.
The estimates are fairly similar in the full sample and the subsample,
which suggests that the presence of ethnic Germans has only little
effect on the estimates of the structural parameters for natives. 

\section{Migrant Effects and Wage Decompositions}\label{sec:experiments}

We proceed to examine actual and counterfactual decompositions of the
wage differential by considering the scenarios of Section \ref{sec:counterfactual.sim}. The
discussion there has highlighted the importance of the productivity
distribution, and we operationalise the decomposition as follows. 

\subsection{Calibration Details}

Our estimation has yielded, given the (estimate of the) actual wage
distribution $G$, the estimated wage offer functions
$w_i^e(p|\hat{\lambda}, \hat{\delta}, \hat{\mu},\hat{\sigma})$. 
Given the Pareto-like productivity distributions, we
calibrate wage offer functions $w_i(p|\hat{\lambda}, \hat{\delta},
\hat{\mu},\hat{\sigma}, \underline{p},\alpha)$ based on Pareto
productivity distributions by minimising the integrated absolute
deviations between $w_i^e(p|.)$ and $w_i(p|.,
\underline{p},\alpha)$. Table \ref{table:Fitting} reports the
calibrated parameters.\footnote{\label{FN:Pareto}
These are also consistent with alternative estimates based on the shapes of 
Figures \ref{fig:young89} - \ref{fig:young7}.
The approximate linearity in the productivity plots suggests a simple
(graphical) estimator of the shape parameter of the Pareto
distributions: use OLS to estimate the regression of  log density on log
productivity (and add 1).}

\begin{table}[htbp]
\small
\begin{center}
\begin{threeparttable}
\caption{Calibrated parameters of the Pareto productivity distribution.} \label{table:Fitting}
\begin{tabular}{r|l|rr|rr|rr}
\hline \hline                                
                                
\multirow{2}{*}{Age Group } &\multirow{2}{*}{Nationality}            &\multicolumn{2}{c|}{Unskilled}&\multicolumn{2}{c|}{Skilled}&\multicolumn{2}{c}{Clerks}\\ \cline{3-8}                               
&            &     $\underline{p}$ &$\alpha$&$\underline{p}$ &$\alpha$&$\underline{p}$ &$\alpha$ \\[.1em]                        
\hline\hline                                
\multirow{2}{*}{25-30}	 &Natives   & 79.789 & 2.511 & 81.282    &  3.172      & 67.677   & 2.449      \\
	                     &Foreigners& 47.632 & 2.205 & 51.010    & 2.146      & 43.053   & 1.468      \\
\multirow{2}{*}{30-40}	 &Natives   & 84.343 & 2.894 &	71.212	  &	 3.076		& 104.849  & 3.096       \\
      	                 &Foreigners& 57.071 & 2.611 &	61.414	  &	 2.661		& 41.818   & 1.463       \\
\multirow{2}{*}{40-55}	 &Natives   & 70.263 & 2.842 &	69.293    &	 3.045      & 72.222   & 3.197       \\
                         &Foreigners& 70.202 & 2.833 &	63.838    &	 2.896		& 34.545   & 1.738      \\
\hline\hline
\end{tabular}
\end{threeparttable}
\end{center}
\end{table}

Figure \ref{fig:sim_1} illustrates these calibrations for young workers in the three
occupations, as well as the counterfactual
experiment of improving the job turnover situation of foreigners by
lowering their job separation rate to those of natives,
$\delta_F \equiv \delta_N$. The  first two columns of the figure show the close match
between $w^e(p)$ (which we have seen before in Figure
\ref{fig:young89}) and $w(p)$. Column three depicts the calibrated wage
offers $w_N(p)$ (solid line) and $w_F(p)$ (dashed line), as well as
the counterfactual $w_F(p|.,\hat{\delta}_N )$ (dotted line). The
reduction in the separation rate for foreigners from $\hat{\delta}_F$
to $\hat{\delta}_N $ `rotates' the wage offer curve up: for lower
productivities, the improvement is negligible, but for very high
productivities foreigners receive wage offers equal to or better than
those for natives. This results in the improvement in the density of
accepted wages depicted in the fourth column of the figure.

\subsection{Results}\label{subsec:experiments}

Tables \ref{table:Decomp1} to \ref{table:Decomp3} report by age
group the average migrant effect (row 1), as well as the results of
the counterfactual experiments which follow the structure of Table
\ref{table:Decomp_SIM}. We can anticipate the qualitative results of
these experiments based on numerical comparative statics exercises
which show (for the set of parameters considered) that wage offers
increase in the job offer arrival rate $\lambda$ as search frictions
decrease, and, contrariwise, decrease in the
job separation rate $\delta$ as search frictions
increase. Of course, as $k=\lambda/\delta \to \infty$ and
  search frictions disappear, by
  eq. (\ref{K_IER_1999}), wage offers converge to the competitive
  wage. Wage offers increase in the mean reservation
wage $\mu$, since by the reservation wage property of job
search only sufficiently high wage offers are accepted out of
unemployment,
but the effect of $\sigma$ is ambiguous. Since we found
that job separation rates for foreigners always exceed those of
natives, setting $\delta_F = \delta_N$ increases their wage offers,
which implies a reduction both of the wage gap and the migrant
effect. Similarly, we found that $\lambda_F > \lambda_N$ (except for
young clerks and service workers), so reducing the foreigners' job
offer arrival rate to that of natives reduces their wage offers, which
implies an increase both in the wage gap and the migrant effect. As
regards reservation wages,  we found that foreigners are on average
less demanding than natives, but the overall effect of the joint
experiment involving $(\mu, \sigma)$ is ambiguous given the ambiguous
effect of $\sigma$. All these qualitative effects are
observed in the results tables (experiments (1)-(4)). The principal
objective of the tables is then to quantify the impacts in order to
understand the principal drivers of the migrant effect.

We comment first on the level of the average migrant effect. Across all age groups,
the absolute migrant effect is the largest for clerks and service
workers, followed by the skilled, and is negligible for the unskilled. In
relative terms, the migrant effect of the skilled accounts for 12-15\% of the average wage gap, and for clerks and service workers for 23-39\%, the average effect being 19.6 \%. Expressed in terms of the average segment-specific wage of natives (see Table \ref{table:DES3}), the migrant effect amounts to 9.2-18.3\% for clerks and service workers, 1.6-7.7\% for skilled workers, 0.7-2.6\% for unskilled workers, and 5.6\% for the full sample. The latter estimates are consistent with
estimates of ``unexplained wage differences'' reported in the
literature for Germany based on standard Oaxaca-Blinder decompositions
(for instance, Lehmer and Ludsteck (2011) report a range from 4 to
17\%) or complementary approaches (\cite{Hirsch&Jahn} report 6\%
while \cite{BartolucciMigration} suggests discrimination effects ranging
between 7 and 17\%). The observed difference between the wage
differential and the migrant effect also implies that the largest part of the
native-migrant wage gap is explained by differences in the
productivity distribution, which is confirmed in experiment (9) by the
drop in the wage gap (which now equals the migrant effect by
construction). Policy interventions that seek to reduce the
productivity gap will thus reduce the wage gap. We turn to the
various experiments, highlighting the role of search frictions.

Consider first the role of the mean reservation wage $\mu$ (experiment
(2)). The gap in mean
reservation wages is the largest for clerks and service
workers whilst the dispersion parameters are fairly similar. Raising
then the foreigners' mean reservation wages shows that the substantial migrant effect for this
occupational group is reduced to between 41\% and 61\% of its
former level. For the
skilled, we only observe a significant gap in mean reservation wages
for the young, and an equalisation of $(\mu, \sigma)$ reduces the
migrant effect to 48\% of its former level. For the other age groups, and for the
unskilled, differences in $\mu$ between natives and foreigners are
either small or negligible, so equalisations have little effect. Once
productivity differences have been eliminated, a comparison between
experiments (9) and (10) shows that for clerks and service workers, the relative
improvement in the migrant effect due to the additional
equalisation of $(\mu, \sigma)$ is slightly larger (the migrant effect
is now between 20\% and 30\% of the level generated in experiment
(9)). Turning to the policy implications, although
foreigners are on average less demanding than natives,
we believe that foreigners' reservation wages should be less a
concern for policy interventions which are
migrant-centred (as emphasised by recent policy debates in the EU,
e.g. \citet[p.~28]{EUReport})
and seek to reduce the migrant effect. Nor would any migrant-targeted
benefit increase be politically feasible in the light of the debate
about welfare magnets.

By the same token, job arrival rates for foreigners typically exceed
those of natives, and thus should equally be of little policy concern. In
fact, the experiments (4) show that reducing this rate to that of natives
only substantially increases the migrant effect for the unskilled in
the two first age groups; for all other groups the induced increase in the
migrant effect is fairly small. This is also in line with the
observation that $\lambda$ falls in age for the unskilled
and skilled.
 
We turn to the remaining frictional parameter, the job separation rate
$\delta$. Recall that foreigners' job separation rates are larger than those
for natives, and sometimes substantially so, and that
  search frictions experienced by them are typically larger than those
  of natives as a higher $\lambda$ cannot compensate for the higher
  $\delta$. Hence  there is scope for migrant-centred policy interventions that
seek to reduce their search frictions, such as 
improving migrants' employment protection. 
This scope, however, decreases in age, as $\delta$
falls in age across all occupational groups. Reducing the 
foreigners' job separation rates to that of natives has the largest
absolute impact for clerks and service workers, followed by the
skilled. For the unskilled, the migrant effect is already fairly small, and an
equalisation of $\delta$ reduces the remainder further. 

\section{Conclusion} \label{conclusion}

The use of the structural empirical general equilibrium search model
with on-the-job search has enabled us to disentangle the role of
various unobservables for the explanation of wage differentials between
migrants and natives. In particular, we have examined differences in
search frictions, reservation wages, and productivities in segments of
the labour market defined by occupation, age, and nationality using a
large scale German administrative dataset. The resulting
decompositions of the actual and counterfactual wage differential
quantify the marginal and joint roles of the various factors.

\section*{Acknowledgements} \label{Acknowledgements}
Financial support from the NORFACE research programme on Migration in
Europe -Social, Economic, Cultural and Policy Dynamics is gratefully
acknowledged. Thanks to Norface conference participants, especially
G. Peri and C. Dustmann for comments, as well as N. Theodoropoulos. We
also wish to thank our three referees whose detailed and very constructive
comments have helped to improve the paper.
\newline \newline
This study uses the factually anonymous regional file of the IAB
Employment Sample (IABS) 1975-2004. Data access was provided via a
Scientific Use File supplied by the Research Data Centre (FDZ) of the
German Federal Employment Agency (BA) at the Institute for Employment
Research (IAB).
\newpage

\begin{landscape}
\begin{figure}[htbp]
\centering
\caption{Calibration, migrant  effects, and wage densities for young workers.}
\label{fig:sim_1}
\includegraphics[scale=0.19] {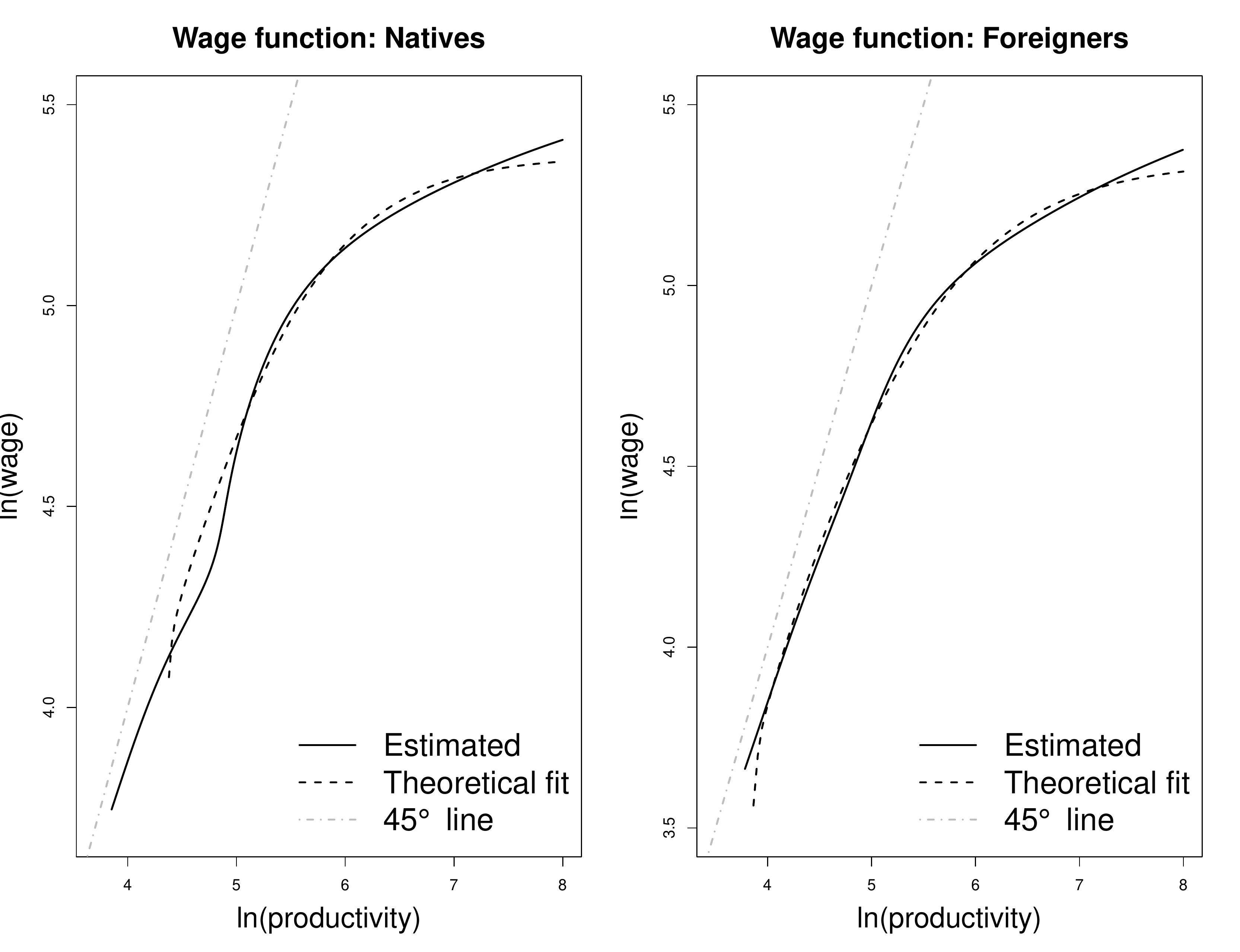}
\includegraphics[scale=0.19] {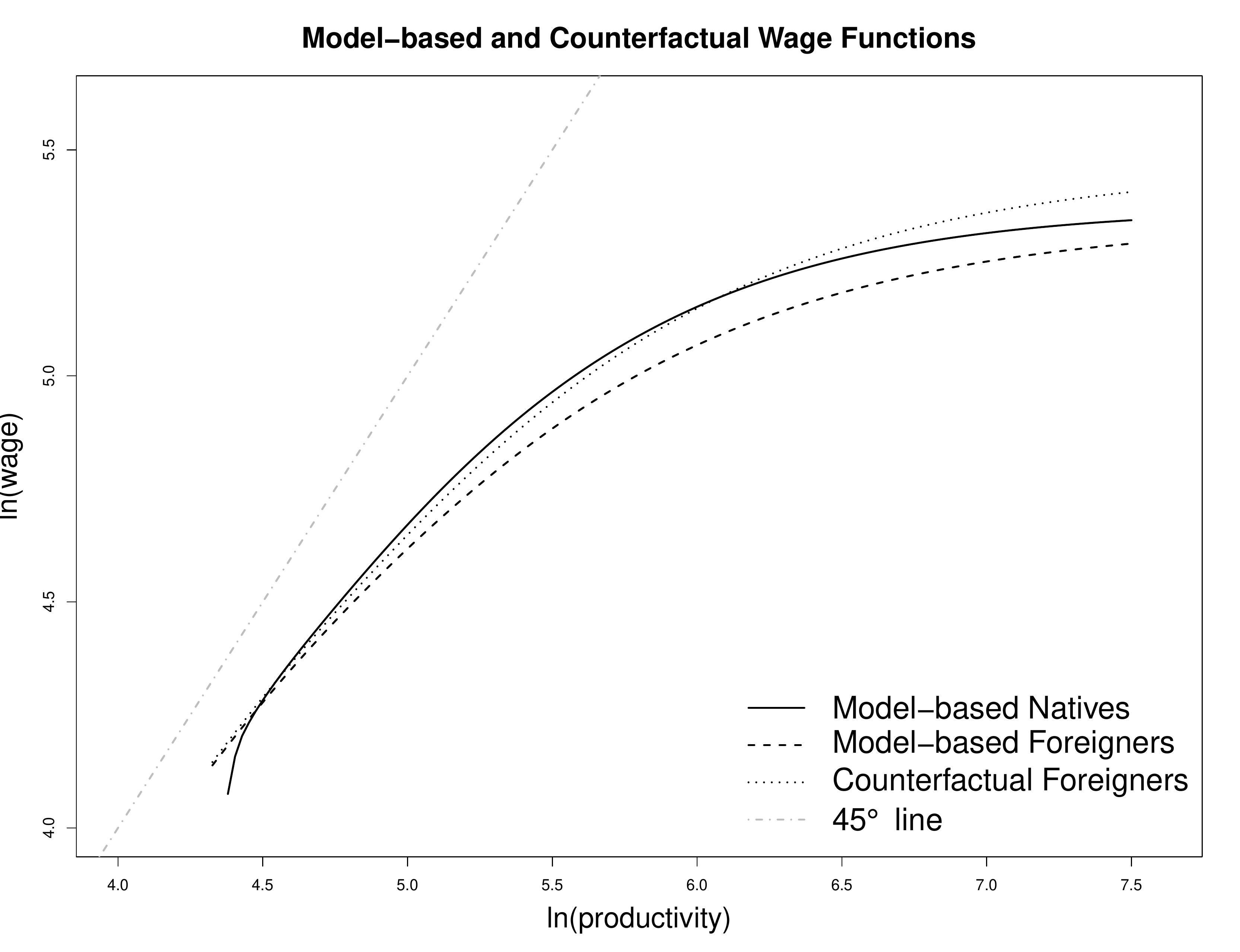} 
\includegraphics[scale=0.19] {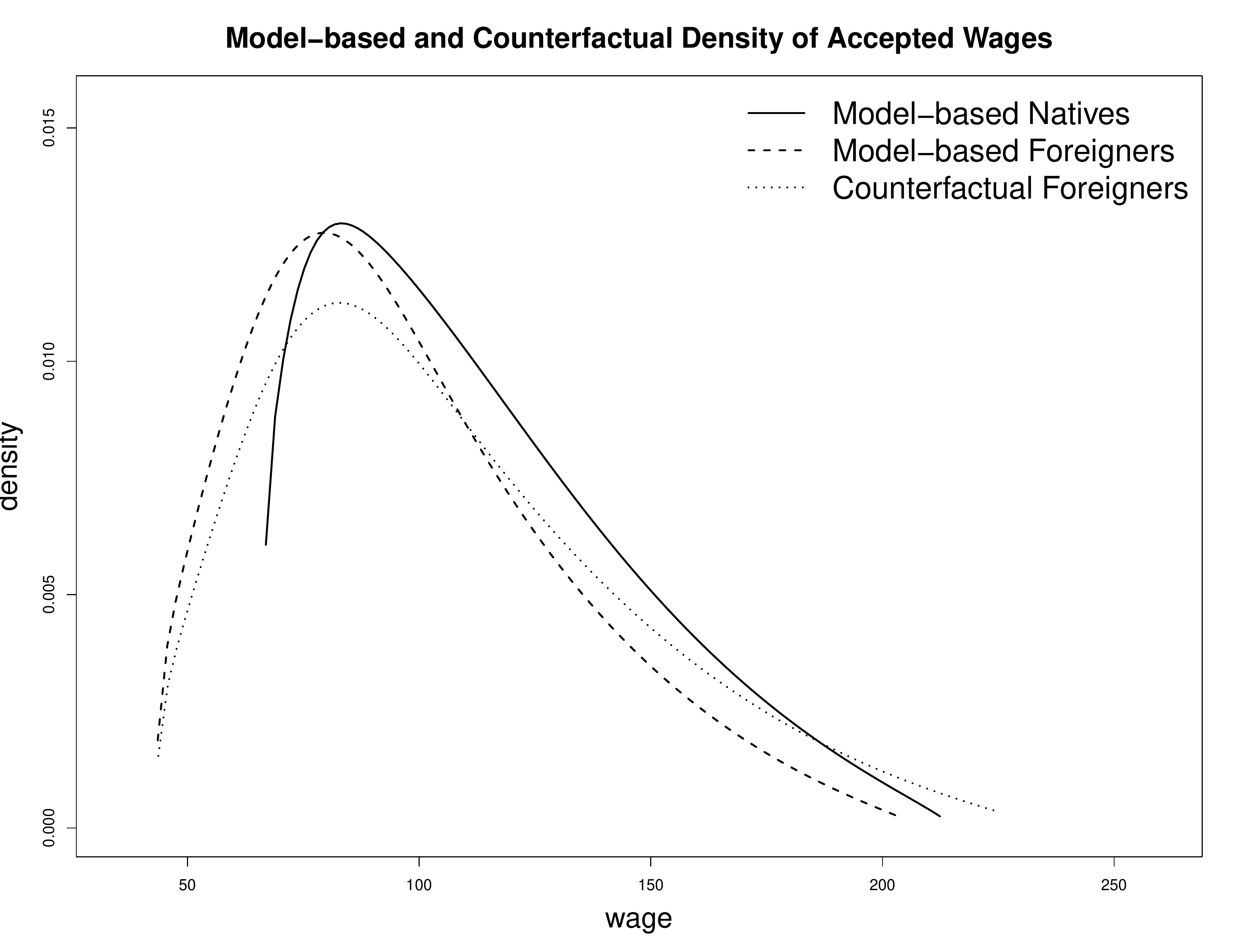}\\
\includegraphics[scale=0.19] {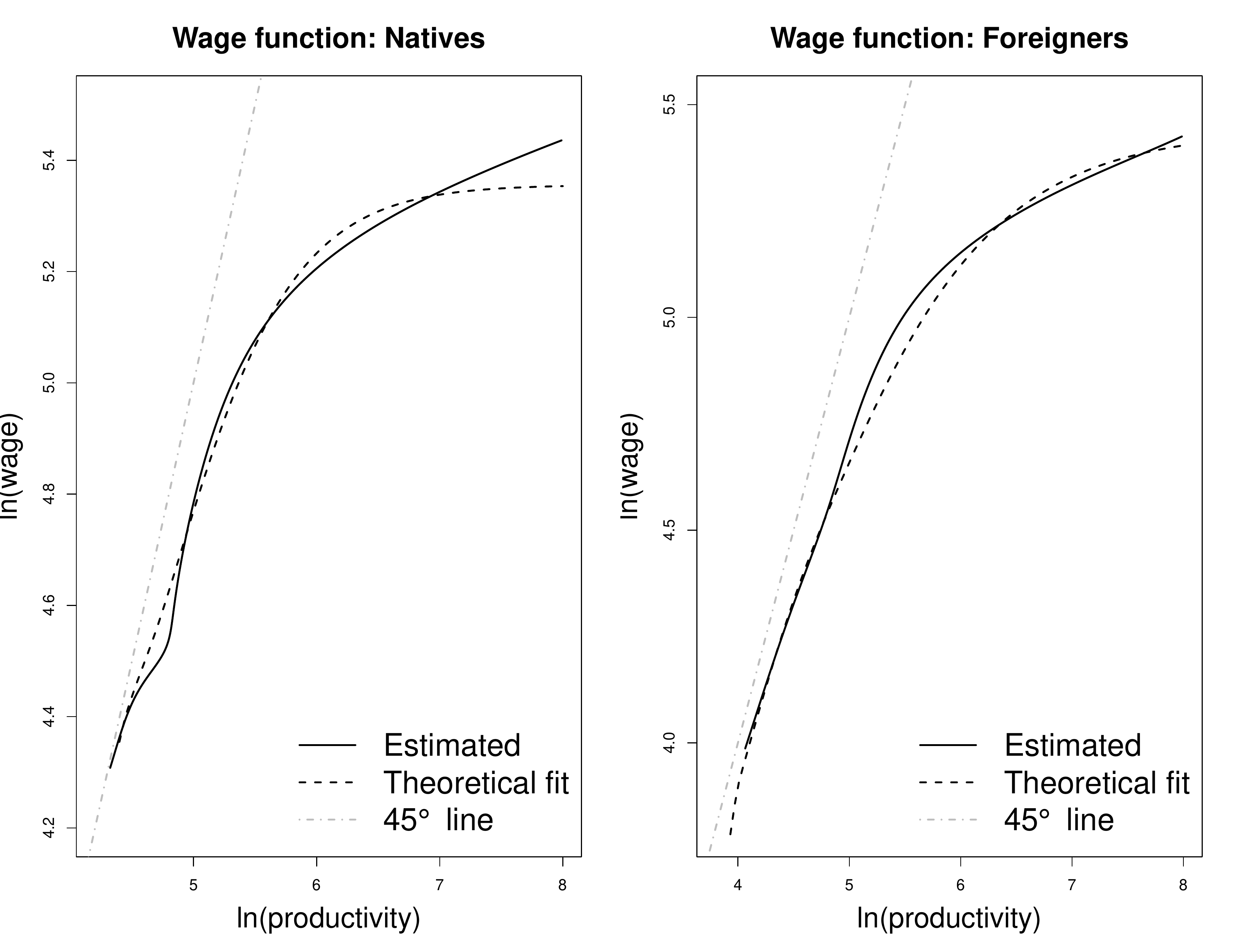}
\includegraphics[scale=0.19] {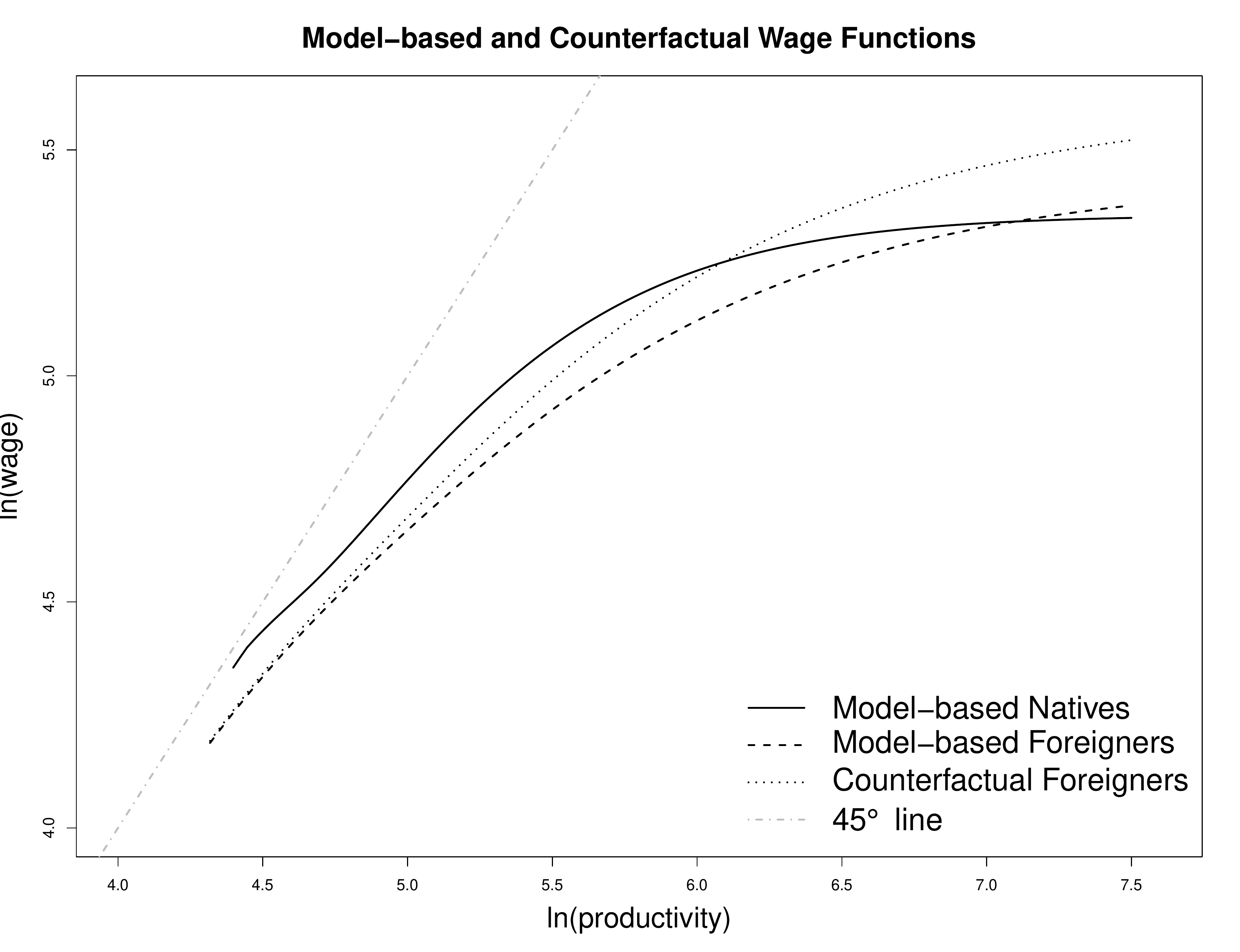} 
\includegraphics[scale=0.19] {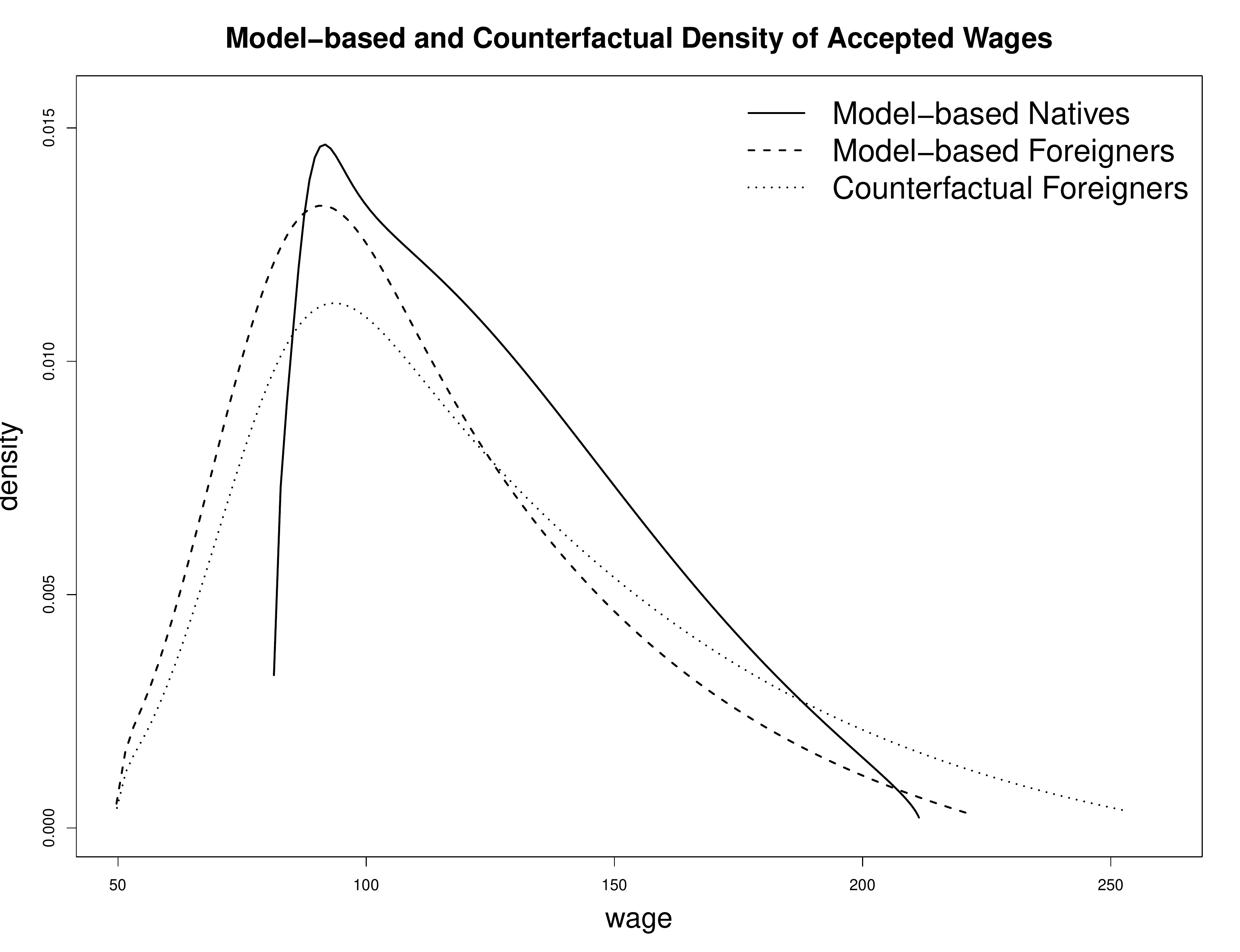}\\
\includegraphics[scale=0.19] {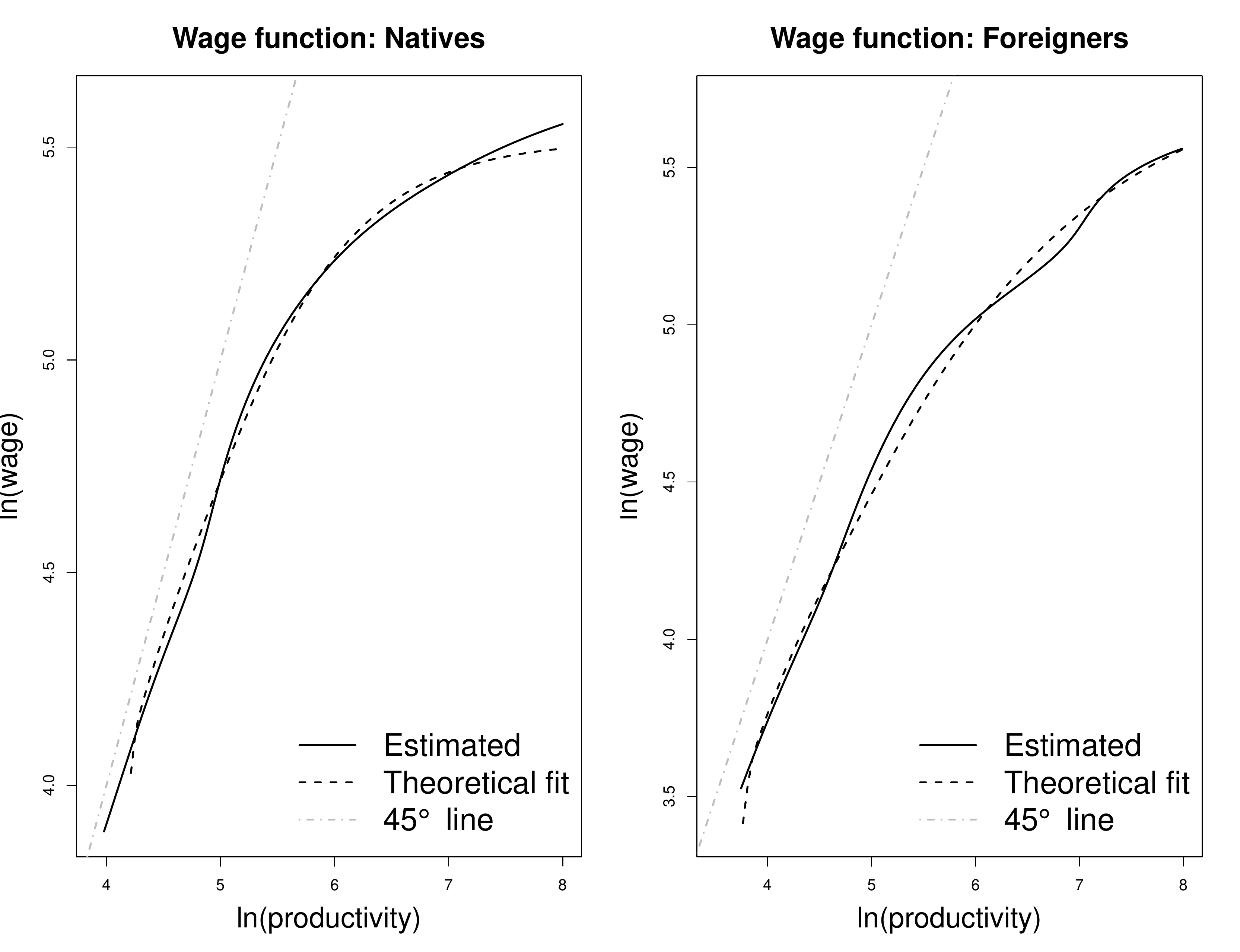}
\includegraphics[scale=0.19] {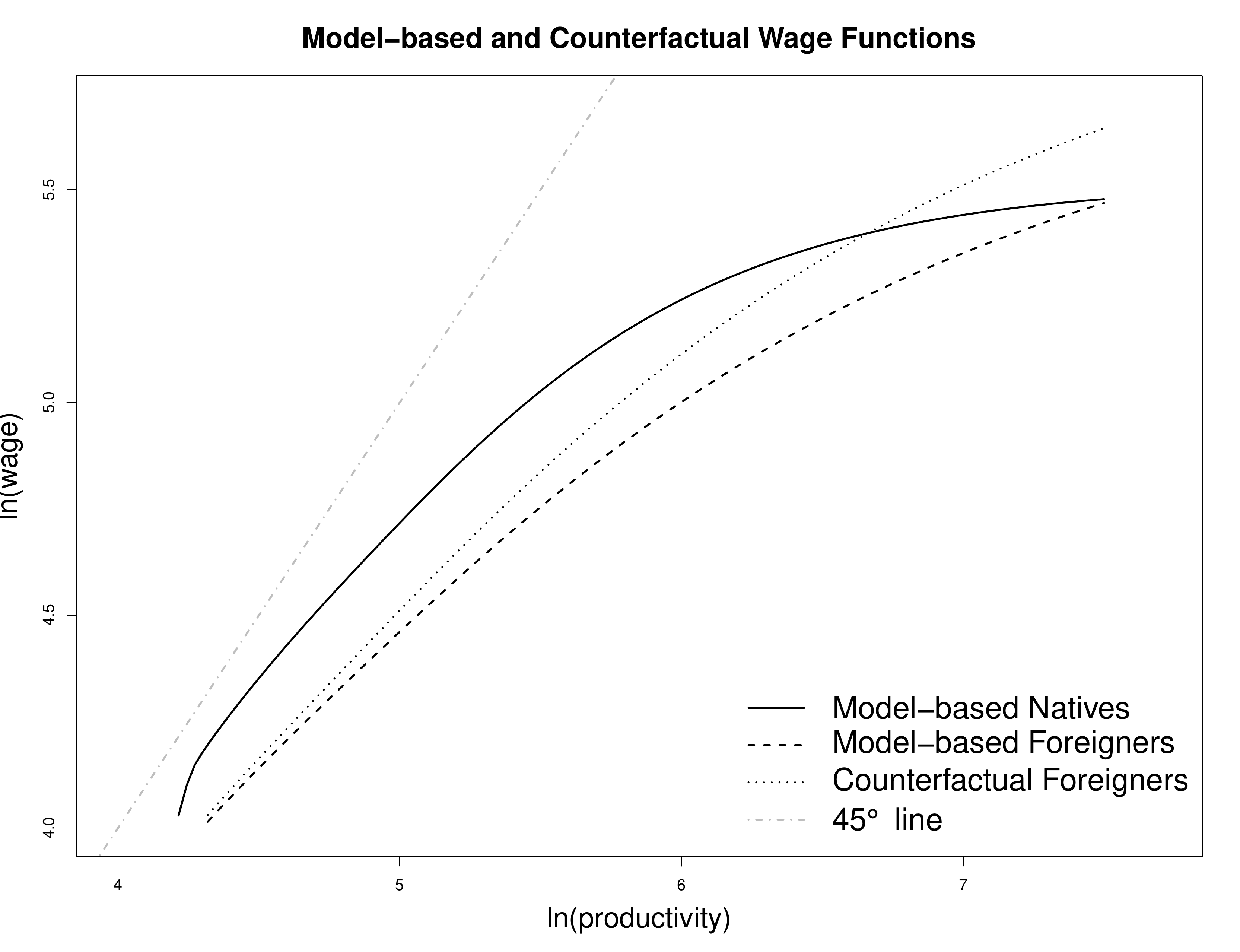} 
\includegraphics[scale=0.19] {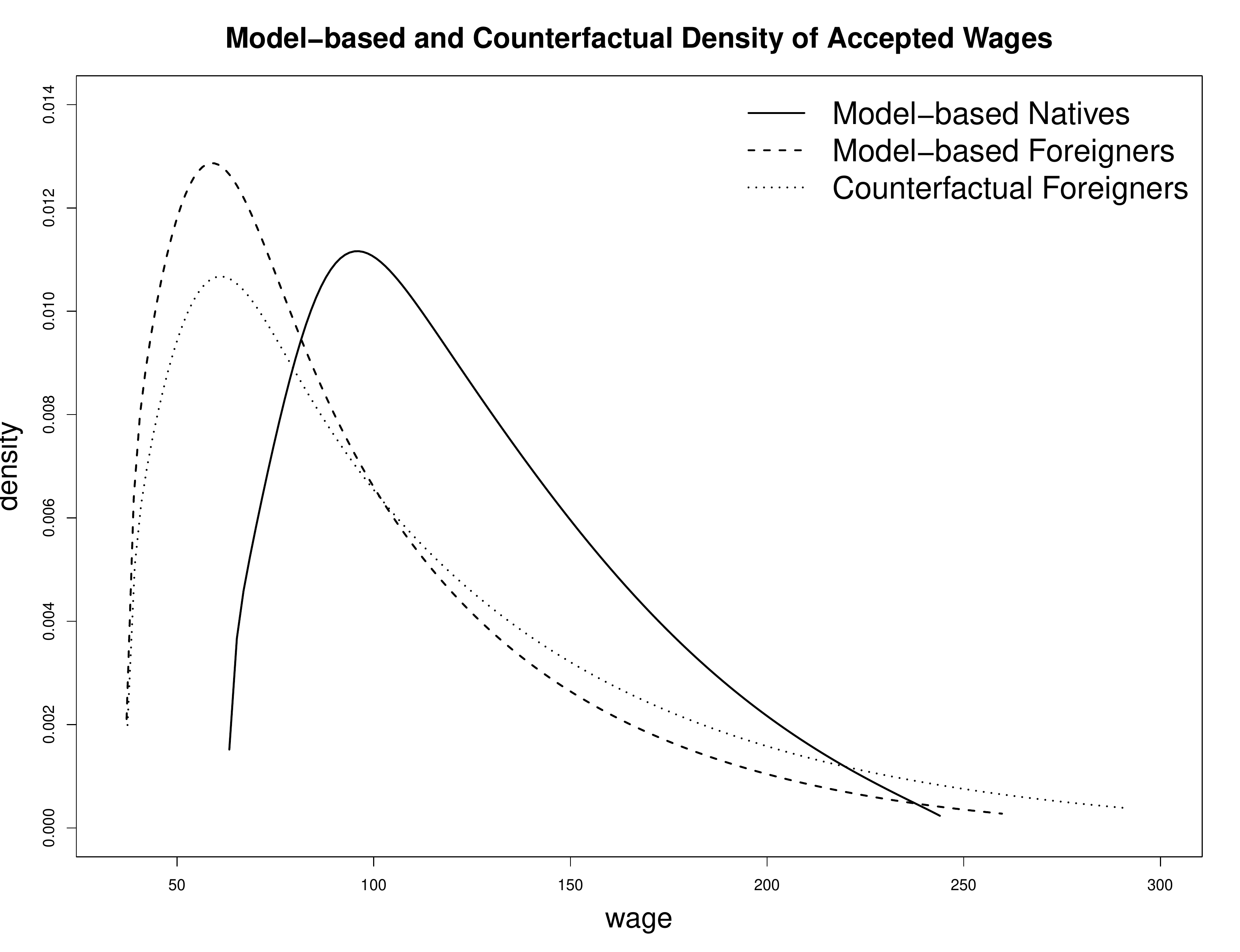}\\
{\footnotesize Notes. Row 1: unskilled blue collar workers. Row 2:
  skilled blue collar workers. Row 3: clerks and lower service workers.}
\end{figure}
\end{landscape}

\begin{landscape}

\begin{table}
\begin{center}
\begin{threeparttable}
\caption{Wage differential decomposition and average migrant effects:
  Ages 25-30.} \label{table:Decomp1}
\begin{tabular}{rll|cc|cc|cc} \hline \hline      
& & & \multicolumn{2}{|c|}{Unskilled}&\multicolumn{2}{c|}{Skilled}&\multicolumn{2}{c}{Clerks}  \\                         
& \multicolumn{1}{l}{Counterfactually } &
\multicolumn{1}{l}{Remaining} & \multicolumn{1}{|l}{Wage}    &
\multicolumn{1}{l}{Migrant} & \multicolumn{1}{|l}{Wage}    &
\multicolumn{1}{l}{Migrant} & \multicolumn{1}{|l}{Wage}    & \multicolumn{1}{l}{Migrant}\\
& \multicolumn{1}{l}{equalised para.} & \multicolumn{1}{l}{differing para.}
& \multicolumn{1}{|l}{differential}    & \multicolumn{1}{l}{effect} &
\multicolumn{1}{|l}{differential}    & \multicolumn{1}{l}{effect} & \multicolumn{1}{|l}{differential}    & \multicolumn{1}{l}{effect}\\
 \hline                               

(1)& & $\underline{p},\alpha,\mu,\sigma,\lambda,\delta$       &   61.717  &   2.779 & 62.463 &   9.560  &  43.426 &  16.929 \\ \hline  
(2)& $\mu,\sigma$ & $\underline{p},\alpha,\lambda,\delta$     &   61.740  &   2.793 & 60.964 &   4.609  &  38.357 &   6.995     \\    
(3)& $\delta$ &$\underline{p},\alpha,\mu,\sigma,\lambda$      &   60.806  &   0.349 & 61.260 &   7.691  &  40.637 &  14.205        \\   
(4)& $\lambda$ & $\underline{p},\alpha,\mu,\sigma,\delta$     &   64.804  &  11.272 & 63.437 &  11.146  &  40.639 &  14.206        \\   
(5)& $\mu,\sigma,\delta$ & $\underline{p},\alpha,\lambda$     &   60.828  &   0.373 & 59.900 &   3.150  &  36.253 &   5.321     \\   
(6)& $\mu,\sigma,\lambda$ & $\underline{p},\alpha,\delta$     &   64.803  &  11.134 & 61.804 &   5.798  &  36.254 &   5.322       \\   
(7)& $\lambda,\delta$ & $\underline{p},\alpha,\mu,\sigma$     &   63.917  &   8.793 & 62.237 &   9.201  &  37.861 &  11.681       \\   
(8)& $\mu,\sigma,\lambda,\delta$ & $\underline{p},\alpha$     &   63.930  &   8.729 & 60.766 &   4.334  &  34.028 &   3.660     \\    
(9)& $\underline{p},\alpha$ & $\mu,\sigma,\lambda,\delta$     &           &  -4.336 &        &   4.142  &         &  11.569    \\    
(10)& $\underline{p},\alpha,\mu,\sigma$ & $\lambda,\delta$    &           &  -4.962 &        &   0.264  &         &   3.501     \\    
(11)& $\underline{p},\alpha,\delta$ & $\mu,\sigma,\lambda$    &           &  -6.137 &        &   2.610  &         &   9.065      \\            
(12)& $\underline{p},\alpha,\lambda$ & $\mu,\sigma,\delta$    &           &   2.989 &        &   5.529  &         &   9.066      \\             
(13)& $\underline{p},\alpha,\mu,\sigma,\delta$ & $\lambda$    &           &  -6.756 &        &  -1.088  &         &   1.710      \\      
(14)& $\underline{p},\alpha,\mu,\sigma,\lambda$ & $\delta$    &           &   2.330 &        &   1.458  &         &   1.711      \\         
(15)& $\underline{p},\alpha, \lambda,\delta$ & $\mu, \sigma$  &           &   0.647 &        &   3.840  &         &   6.812     \\        
\hline\hline
\end{tabular}
\footnotesize Notes: Based on the decomposition of equation (\ref{counterdecomp1}). Rows 9$+$: the wage differential
  equals the migrant effect because the productivity distributions are
  the same. The parameter estimates are reported in Tables \ref{table:UBCW_A_1}- \ref{table:UBCW_A_3}.
\end{threeparttable}
\end{center}
\end{table}

\begin{table}
\begin{center}
\begin{threeparttable}
\caption{Wage differential decomposition and average migrant effects:
  Ages 30-40.} \label{table:Decomp2}
\begin{tabular}{rll|cc|cc|cc} \hline \hline      
& & & \multicolumn{2}{|c|}{Unskilled}&\multicolumn{2}{c|}{Skilled}&\multicolumn{2}{c}{Clerks}  \\                         
& \multicolumn{1}{l}{Counterfactually } &
\multicolumn{1}{l}{Remaining} & \multicolumn{1}{|l}{Wage}    &
\multicolumn{1}{l}{Migrant} & \multicolumn{1}{|l}{Wage}    &
\multicolumn{1}{l}{Migrant} & \multicolumn{1}{|l}{Wage}    & \multicolumn{1}{l}{Migrant}\\
& \multicolumn{1}{l}{equalised para.} & \multicolumn{1}{l}{differing para.}
& \multicolumn{1}{|l}{differential}    & \multicolumn{1}{l}{effect} &
\multicolumn{1}{|l}{differential}    & \multicolumn{1}{l}{effect} & \multicolumn{1}{|l}{differential}    & \multicolumn{1}{l}{effect}\\
 \hline                               

(1)& & $\underline{p},\alpha,\mu,\sigma,\lambda,\delta$       &  59.484  &  0.854 & 29.662  &  3.904 &  89.489 &  28.642  \\ \hline  
(2)& $\mu,\sigma$ & $\underline{p},\alpha,\lambda,\delta$     &  59.699  &  1.456 & 28.539  &  2.228 &  87.086 &  17.559    \\    
(3)& $\delta$ &$\underline{p},\alpha,\mu,\sigma,\lambda$      &  57.916  & -2.964 & 27.915  &  1.956 &  84.023 &  19.278       \\   
(4)& $\lambda$ & $\underline{p},\alpha,\mu,\sigma,\delta$     &  62.756  &  9.059 & 30.355  &  4.696 &  91.897 &  33.358       \\   
(5)& $\mu,\sigma,\delta$ & $\underline{p},\alpha,\lambda$     &  58.089  & -2.472 & 26.937  &  0.476 &  82.447 &  11.304    \\   
(6)& $\mu,\sigma,\lambda$ & $\underline{p},\alpha,\delta$     &  63.106  & 10.004 & 29.159  &  2.922 &  88.962 &  20.378      \\   
(7)& $\lambda,\delta$ & $\underline{p},\alpha,\mu,\sigma$     &  60.872  &  4.298 & 28.503  &  2.603 &  86.404 &  23.116      \\   
(8)& $\mu,\sigma,\lambda,\delta$ & $\underline{p},\alpha$     &  61.136  &  5.027 & 27.482  &  1.065 &  84.526 &  13.974    \\    
(9)& $\underline{p},\alpha$ & $\mu,\sigma,\lambda,\delta$     &          & -3.091 &         &  2.547 &         &  14.515   \\    
(10)& $\underline{p},\alpha,\mu,\sigma$ & $\lambda,\delta$    &          & -2.585 &         &  1.097 &         &   3.011    \\    
(11)& $\underline{p},\alpha,\delta$ & $\mu,\sigma,\lambda$    &          & -5.709 &         &  0.758 &         &   8.259     \\       
(12)& $\underline{p},\alpha,\lambda$ & $\mu,\sigma,\delta$    &          &  3.264 &         &  3.286 &         &  18.087     \\       
(13)& $\underline{p},\alpha,\mu,\sigma,\delta$ & $\lambda$    &          & -5.249 &         & -0.549 &         &  -2.022     \\      
(14)& $\underline{p},\alpha,\mu,\sigma,\lambda$ & $\delta$    &          &  3.938 &         &  1.763 &         &   5.737     \\       
(15)& $\underline{p},\alpha, \lambda,\delta$ & $\mu, \sigma$  &          & -0.563 &         &  1.349 &         &  10.719    \\ 
\hline\hline
\end{tabular}
\footnotesize Notes: As for Table \ref{table:Decomp1}.
\end{threeparttable}
\end{center}
\end{table}

\begin{table}
\begin{center}
\begin{threeparttable}
\caption{Wage differential decomposition and average migrant effects:
  Ages 40-55.} \label{table:Decomp3}
\begin{tabular}{rll|cc|cc|cc} \hline \hline      
& & & \multicolumn{2}{|c|}{Unskilled}&\multicolumn{2}{c|}{Skilled}&\multicolumn{2}{c}{Clerks}  \\                         
& \multicolumn{1}{l}{Counterfactually } &
\multicolumn{1}{l}{Remaining} & \multicolumn{1}{|l}{Wage}    &
\multicolumn{1}{l}{Migrant} & \multicolumn{1}{|l}{Wage}    &
\multicolumn{1}{l}{Migrant} & \multicolumn{1}{|l}{Wage}    & \multicolumn{1}{l}{Migrant}\\
& \multicolumn{1}{l}{equalised para.} & \multicolumn{1}{l}{differing para.}
& \multicolumn{1}{|l}{differential}    & \multicolumn{1}{l}{effect} &
\multicolumn{1}{|l}{differential}    & \multicolumn{1}{l}{effect} & \multicolumn{1}{|l}{differential}    & \multicolumn{1}{l}{effect}\\
 \hline                               
(1)& & $\underline{p},\alpha,\mu,\sigma,\lambda,\delta$       &  2.817 &  2.723  & 19.784  &  2.327 &  64.712 &  14.591\\ \hline  
(2)& $\mu,\sigma$ & $\underline{p},\alpha,\lambda,\delta$     &  1.728 &  1.631  & 18.883  &  1.172 &  62.721 &   6.298    \\    
(3)& $\delta$ &$\underline{p},\alpha,\mu,\sigma,\lambda$      &  1.145 &  1.056  & 18.971  &  1.404 &  62.127 &  11.248       \\   
(4)& $\lambda$ & $\underline{p},\alpha,\mu,\sigma,\delta$     &  2.788 &  2.694  & 20.030  &  2.608 &  65.002 &  15.012       \\   
(5)& $\mu,\sigma,\delta$ & $\underline{p},\alpha,\lambda$     &  0.125 &  0.032  & 18.114  &  0.304 &  60.519 &   4.125    \\   
(6)& $\mu,\sigma,\lambda$ & $\underline{p},\alpha,\delta$     &  1.700 &  1.603  & 19.113  &  1.433 &  62.954 &   6.553      \\   
(7)& $\lambda,\delta$ & $\underline{p},\alpha,\mu,\sigma$     &  1.120 &  1.030  & 19.193  &  1.655 &  62.374 &  11.533      \\   
(8)& $\mu,\sigma,\lambda,\delta$ & $\underline{p},\alpha$     &  0.100 &  0.008  & 18.325  &  0.541 &  60.737 &   4.321    \\    
(9)& $\underline{p},\alpha$ & $\mu,\sigma,\lambda,\delta$     &        &  2.716  &         &  1.699 &         &   7.375   \\    
(10)& $\underline{p},\alpha,\mu,\sigma$ & $\lambda,\delta$    &        &  1.623  &         &  0.601 &         &   1.676    \\    
(11)& $\underline{p},\alpha,\delta$ & $\mu,\sigma,\lambda$    &        &  1.049  &         &  0.828 &         &   5.173     \\       
(12)& $\underline{p},\alpha,\lambda$ & $\mu,\sigma,\delta$    &        &  2.687  &         &  1.966 &         &   7.660     \\       
(13)& $\underline{p},\alpha,\mu,\sigma,\delta$ & $\lambda$    &        &  0.025  &         & -0.225 &         &  -0.160     \\      
(14)& $\underline{p},\alpha,\mu,\sigma,\lambda$ & $\delta$    &        &  1.595  &         &  0.851 &         &   1.905     \\       
(15)& $\underline{p},\alpha, \lambda,\delta$ & $\mu, \sigma$  &        &  1.023  &         &  1.064 &         &   5.359    \\      
\hline\hline
\end{tabular}
\footnotesize Notes: As for Table \ref{table:Decomp1}.
\end{threeparttable}
\end{center}
\end{table}

\end{landscape}

\appendix 

\section{Data Appendix: Variable Description}

Our sample only includes full-time working men aged
25-55 years old residing in West Germany. In what follows, we describe how we
construct the key variables used in our empirical analysis.

\noindent \textbf{Age:} The age variable is constructed using information on the date of
birth and the year in which the spell took place. Date of birth is not
available for individuals who were under 16 years old at their first observed
spell or over 65 years old at their last observed spell. In such cases, we
assume that workers were 15 years old at their first spell and 67 years old
at their last spell.

\noindent \textbf{Labour Market Status:} The information provided in the data set are sufficient to distinguish between three labour market states: employed, recipient of transfer payments, and out of sample. In our analysis, we employ the broad definition of unemployment, as proposed by Fitzenberger and Wilke (2010), and assume that unemployment is proxied by non-employment. Using this definition of unemployment, we only consider two labour market states (employment and unemployment) since being out of sample is equivalent to being unemployed. However, this strategy may lead to the mis-classification of non-participants as unemployed: for example, an individual that had an employment spell in her late teens, subsequently went to university, and reappeared in the sample in her late twenties would be classified as unemployed despite the fact that she was not in the labour market. To correct for this problem, individuals that are out of sample are only classified as unemployed if their out of sample duration does not exceed the average duration of transfer payment recipients' spells.

\noindent \textbf{Spells:} Due to the annual reporting system, all spells have a maximum
duration of one year. We merge all consecutive annual spells during which the
individual does not experience a change in her labour market status, i.e. she
either remains unemployed or employed with the same employer. We use
firm-identifiers included in the dataset to determine when a worker changes
employers. The new merged spells record the start date, the end date, the
duration of the spell, the employment status, the average wage under the
same employer, and the transition experienced by the individual
(job-to-unemployment, job-to-job, unemployment-to-job).

\noindent \textbf{Wages:} The dataset reports gross daily wages and does not provide
information on hours worked. We therefore exclude part-time employees,
trainees, interns, and at-home workers from the sample since the wage
information is not comparable for these groups. Wages are truncated at the
social security contributions threshold (DM10) and censored at the social
security contributions ceiling (DM300). For workers with wages below the
social security contribution threshold, we use wages of adjacent employment
spells. Wage censoring is not pronounced as the social security contributions
ceiling is not binding in our sample as we focus on low-wage workers who are not
likely to earn wages in excess of this upper bound. 

Since the focus of our analysis is the transitions experienced by
workers in the early 1990s, all wages are reported in DM and adjusted
to real 1995 prices using the  German Consumer Price Index. For all
individuals who experience wage variation during employment spells, we
compute the average per period wage of each worker under the same
employer. 

\noindent \textbf{Occupation:} The dataset includes extensive information on occupations
(three-digit codes), which is used to classify individuals to 10 major groups
based on the International Standard Classification of Occupations (ISCO-88).
Exploiting the detailed
index of occupational titles of the ISCO-88, we are able to map the code list
from the Federal Employment Service of Germany included in the IABS into the following ISCO-88
major groups: (1) Legislators, Senior Officials, and Managers; (2)
Professionals; (3) Technicians and Associate Professionals; (4)
Clerks; (5) Service Workers and Shop \& Market Sales Workers; (6)
Skilled Agricultural and Fishery Workers; (7) Craft and Related Trades
Workers; (8) Plant and Machine Operators and Assemblers; (9)
Elementary Occupations; (10) Armed Forces.

We restrict attention to low- and middle-skill occupations, where the
concentration of foreigners is higher. Specifically, we consider three
occupational groups that are defined as follows: (1) Unskilled
blue-collar workers, which includes individuals classified in the
ISCO-88 major groups 8 and 9; (2) Skilled blue-collar workers, which
includes individuals classified in the ISCO-88 major group 7; (3)
Clerks \& low-service workers, which includes individuals classified
in the ISCO-88 major groups 4 and 5.

\section{Estimation: A Validation Exercise}\label{sec:validation}

Given the complexity of both the model and the estimating equations,
it is of interest to test their performance in a simulation
exercise. In this appendix, we carry out such a validation exercise. 

The data generating process uses the parametrisations discussed above:
arrival of job offers and separations follow Poisson processes,
and the reservation wage distribution is normal. The particular
calibration, given in Table \ref{table:Simulation_Natives},
distinguishes between the segments for natives (subscripted N) and
immigrants (subscripted F), and uses values
similar to those encountered in our data. We
also need to stipulate either a realised wage distribution $G$, or a
productivity distribution $\Gamma$. Since we observe wages but not
productivities in our data, we specify a productivity distribution here in order
to verify that the model-implied wage distributions ``look realistic''
(i.e. share the principal features of real wage distributions). Since
the empirical results suggest that
productivities are Pareto-like, we assume this explicitly here: $\Gamma_F
(p)=1-(\underline{p}_F/p)^{\alpha}$ and $\Gamma_N
(p)=1-(\underline{p}_N/p)^{\alpha}$ with $\alpha=2.1$,
$\underline{p}_F=40$, and $\underline{p}_N=50$. Hence the productivity
distribution in the segment for natives first order stochastically
dominates that of migrants. We also compute the
model-implied unemployment rate $u$. Using this
Data Generating Process (DGP), we draw 400 samples of 2000
observations each and estimate the model by maximum likelihood. 

\begin{table}[htbp]
\begin{center}
\addtolength{\tabcolsep}{-1.5pt}
\begin{threeparttable}
\caption{Natives and immigrants: DGP and parameter estimates.} \label{table:Simulation_Natives}
\begin{tabular}{ccccccccccc}
\hline \hline
& $\mu_N$ & $\mu_F$ & $\sigma_N$ & $\sigma_F$ & $\lambda_N$ & $\lambda_F$ & $\delta_N$ & $\delta_F$ & $u_N$ & $u_F$ \\
True Value & 60 & 45 & 10 & 10  & .07 & .13 & .005 & .016 & .1214 & .1838 \\ \hline
Mean & 56.23 &40.88   &8.61 &10.18  & .0887 & .1181 &.0050 & .0173   &.1145 & .1822 \\
Median & 56.33 & 40.96   & 8.43  & 10.17  & .0835 & .1136  & .0050 & .0173   &.1142 & .1819  \\
2.5 perc. & 53.46  &  36.62 &   5.63 & 6.86 & .0566  & .0939 & .0047 & .0164 & .1053 & .1711 \\
97.5 perc. &  59.88 & 45.21 &  12.38 & 13.62 &  .1403 & .1671  & .0053 & .0181 & .1246 & .1935 \\
\hline\hline
\end{tabular}
\end{threeparttable}
\end{center}
\end{table}

Section \ref{sec:decomposition} has considered the economic
implications of the estimation results. Here, the main focus is on the
quality of the estimates.
Table \ref{table:Simulation_Natives} reports the results. 
All structural parameters are
estimated well as the true values are included in the 95\%
bootstrap confidence intervals (the table reports the 2.5 and 97.5 \%
confidence limits). The means of the job turnover
parameters are particularly well estimated. The mean of the
reservation wage distribution $H$ is somewhat below the true value; this
underestimate is perhaps not too surprising since the model effectively only
considers the right tail of $H$ (i.e. reservation wages $b$ that
satisfy $b > \underbar{w}$). The predicted unemployment rate is also
very close to the theoretical value. 

Figure \ref{fig:Migrant_effect} above has depicted the implied wage offers as a
function of productivities\footnote{The computation of the wage offer curves for the
  validation exercise based on a given productivity distribution
  $\Gamma$ is more involved than in our empirical analysis below. In
  the latter case, given the estimates of the structural parameters
  and the wage density, $F(w)$ follows straightforwardly from equation
  (\ref{SS_A1}) and the productivity values follow from
  (\ref{Prod_A}). In the former case, $F\left( w\right) =\Gamma \left(
    K^{-1}\left( w\right)\right)$, and $K(p)$ in (\ref{K_IER_1999}),
  defined implicitly, is estimated progressively: starting from
  $\underline{p}$, $p$ is incremented by a small step $\varepsilon_p$, and
  $K(p+\varepsilon_p)$ is found through a local search based on
  (\ref{K_IER_1999}), whence $p+2 \varepsilon_p$ is considered. The
  confidence bands are computed pointwise, and simply determined by
  the relevant tail quantiles of the bootstrap distribution.} 
while Figure \ref{fig:sim_1} has depicted the skewed densities of
realised wages, which do have a shape often
encountered in empirical work (see e.g. Figure \ref{fig:wage_density}). 

\newpage

\bibliography{Master2}

\begin{thebibliography}{}

\bibitem[\protect\citeauthoryear{Albrecht and Axell}{Albrecht and
  Axell}{1984}]{Albrecht&Axell}
Albrecht, J.~W. and B.~Axell (1984).
\newblock An equilibrium model of search unemployment.
\newblock {\em Journal of Political Economy\/}~{\em 92\/}(5), 824--40.

\bibitem[\protect\citeauthoryear{Aydemir and Skuterud}{Aydemir and
  Skuterud}{2008}]{Aydemir&Skut}
Aydemir, A. and M.~Skuterud (2008).
\newblock The immigrant wage differential within and across establishments.
\newblock {\em Industrial and Labor Relations Review\/}~{\em 61\/}(3),
  334--352.

\bibitem[\protect\citeauthoryear{Barth and Dale-Olsen}{Barth and
  Dale-Olsen}{2009}]{Barth&Dale-Olsen}
Barth, E. and H.~Dale-Olsen (2009).
\newblock Monopsonistic discrimination, worker turnover, and the gender wage
  gap.
\newblock {\em Labour Economics\/}~{\em 16\/}(5), 589--597.

\bibitem[\protect\citeauthoryear{Bartolucci}{Bartolucci}{2013a}]{BartolucciGen%
der}
Bartolucci, C. (2013a).
\newblock Gender wage gaps reconsidered: {A} structural approach using matched
  employer-employee data.
\newblock {\em Journal of Human Resources\/}, to appear.

\bibitem[\protect\citeauthoryear{Bartolucci}{Bartolucci}{2013b}]{BartolucciMig%
ration}
Bartolucci, C. (2013b).
\newblock Understanding the native-immigrant wage gap using matched
  employer-employee data. {E}vidence from {G}ermany.
\newblock {\em Industrial and Labor Relations Review\/}, to appear.

\bibitem[\protect\citeauthoryear{Bender, Haas, and Klose}{Bender
  et~al.}{2000}]{Benderetal}
Bender, S., A.~Haas, and C.~Klose (2000).
\newblock The {IAB} employment subsample 1975-1995.
\newblock {\em Schmollers Jahrbuch\/}~{\em 120\/}(4), 649--662.

\bibitem[\protect\citeauthoryear{Bergemann, Caliendo, van~den Berg, and
  Zimmermann}{Bergemann et~al.}{2011}]{Bergemannetal}
Bergemann, A., M.~Caliendo, G.~J. van~den Berg, and K.~F. Zimmermann (2011).
\newblock The threat effect of participation in active labor market programs on
  job search behavior of migrants in {G}ermany.
\newblock {\em International Journal of Manpower\/}~{\em 32\/}(7), 777--795.

\bibitem[\protect\citeauthoryear{Bontemps, Robin, and Van~den Berg}{Bontemps
  et~al.}{1999}]{Bontempsetal1999}
Bontemps, C., J.-M. Robin, and G.~J. Van~den Berg (1999).
\newblock An empirical equilibrium job search model with search on the job and
  heterogeneous workers and firms.
\newblock {\em International Economic Review\/}~{\em 40\/}(4), 1039--74.

\bibitem[\protect\citeauthoryear{Bowlus}{Bowlus}{1997}]{Bowlus1997}
Bowlus, A.~J. (1997).
\newblock A search interpretation of male-female wage differentials.
\newblock {\em Journal of Labor Economics\/}~{\em 15\/}(4), 625--57.

\bibitem[\protect\citeauthoryear{Bowlus and Eckstein}{Bowlus and
  Eckstein}{2002}]{Bowlus&Eckstein}
Bowlus, A.~J. and Z.~Eckstein (2002).
\newblock Discrimination and skill differences in an equilibrium search model.
\newblock {\em International Economic Review\/}~{\em 43\/}(4), 1309--1345.

\bibitem[\protect\citeauthoryear{Br\"{u}cker and Jahn}{Br\"{u}cker and
  Jahn}{2011}]{Bruecker&Jahn}
Br\"{u}cker, H. and E.~J. Jahn (2011).
\newblock Migration and wage-setting: {R}eassessing the labor market effects of
  migration.
\newblock {\em Scandinavian Journal of Economics\/}~{\em 113}, 286--317.

\bibitem[\protect\citeauthoryear{Burdett and Mortensen}{Burdett and
  Mortensen}{1998}]{Burdett&Mortensen}
Burdett, K. and D.~T. Mortensen (1998).
\newblock Wage differentials, employer size, and unemployment.
\newblock {\em International Economic Review\/}~{\em 39\/}(2), 257--73.

\bibitem[\protect\citeauthoryear{Cahuc, Postel-Vinay, and Robin}{Cahuc
  et~al.}{2006}]{Cahucetal2006}
Cahuc, P., F.~Postel-Vinay, and J.-M. Robin (2006).
\newblock Wage bargaining with on-the-job search: {T}heory and evidence.
\newblock {\em Econometrica\/}~{\em 74\/}(2), 323--364.

\bibitem[\protect\citeauthoryear{Christensen, Lentz, Mortensen, Neumann, and
  Werwatz}{Christensen et~al.}{2005}]{Christensenetal}
Christensen, B.~J., R.~Lentz, D.~T. Mortensen, G.~R. Neumann, and A.~Werwatz
  (2005).
\newblock On-the-job search and the wage distribution.
\newblock {\em Journal of Labor Economics\/}~{\em 23\/}(1), 31--58.

\bibitem[\protect\citeauthoryear{D'Amuri, Ottaviano, and Peri}{D'Amuri
  et~al.}{2010}]{DAmurietal}
D'Amuri, F., G.~I. Ottaviano, and G.~Peri (2010).
\newblock The labor market impact of immigration in {W}estern {G}ermany in the
  1990s.
\newblock {\em European Economic Review\/}~{\em 54\/}(4), 550--570.

\bibitem[\protect\citeauthoryear{de~Matos}{de~Matos}{2012}]{Damas}
de~Matos, A.~D. (2012).
\newblock The careers of immigrants.
\newblock CEP Discussion Papers 1171, Centre for Economic Performance, LSE.

\bibitem[\protect\citeauthoryear{Dustmann, Glitz, and Vogel}{Dustmann
  et~al.}{2010}]{DustmannetalEER}
Dustmann, C., A.~Glitz, and T.~Vogel (2010).
\newblock Employment, wages, and the economic cycle: {D}ifferences between
  immigrants and natives.
\newblock {\em European Economic Review\/}~{\em 54\/}(1), 1--17.

\bibitem[\protect\citeauthoryear{Dustmann and Theodoropoulos}{Dustmann and
  Theodoropoulos}{2010}]{Dustmann&Theo}
Dustmann, C. and N.~Theodoropoulos (2010).
\newblock Ethnic minority immigrants and their children in {B}ritain.
\newblock {\em Oxford Economic Papers\/}~{\em 62\/}(2), 209--233.

\bibitem[\protect\citeauthoryear{Eckstein and van~den Berg}{Eckstein and
  van~den Berg}{2007}]{Eckstein&VanDenBerg}
Eckstein, Z. and G.~J. van~den Berg (2007).
\newblock Empirical labor search: A survey.
\newblock {\em Journal of Econometrics\/}~{\em 136\/}(2), 531--564.

\bibitem[\protect\citeauthoryear{EUCommission}{EUCommission}{2012}]{EUReport}
EUCommission (2012).
\newblock {FP7-SSH-2013} work programme.
\newblock Technical Report C4536.

\bibitem[\protect\citeauthoryear{Fitzenberger, Osikominu, and
  Voelter}{Fitzenberger et~al.}{2006}]{Fitzetal}
Fitzenberger, B., A.~Osikominu, and R.~Voelter (2006).
\newblock Imputation rules to improve the education variable in the {IAB}
  employment subsample.
\newblock {\em Schmollers Jahrbuch\/}~{\em 126\/}(3), 405--436.

\bibitem[\protect\citeauthoryear{Fitzenberger and Wilke}{Fitzenberger and
  Wilke}{2010}]{Fitz&Wilke}
Fitzenberger, B. and R.~A. Wilke (2010).
\newblock Unemployment durations in {W}est {G}ermany before and after the
  reform of the unemployment compensation system during the 1980s.
\newblock {\em German Economic Review\/}~{\em 11}, 336--366.

\bibitem[\protect\citeauthoryear{Flabbi}{Flabbi}{2010}]{FlabbiIER}
Flabbi, L. (2010).
\newblock Gender discrimination estimation in a search model with matching and
  bargaining.
\newblock {\em International Economic Review\/}~{\em 51\/}(3), 745--783.

\bibitem[\protect\citeauthoryear{Glitz}{Glitz}{2012}]{Glitz2012}
Glitz, A. (2012).
\newblock Ethnic segregation in {G}ermany.
\newblock IZA Discussion Papers 6841, Institute for the Study of Labor (IZA).

\bibitem[\protect\citeauthoryear{Gundel and Peters}{Gundel and
  Peters}{2008}]{Gundel&Peters}
Gundel, S. and H.~Peters (2008).
\newblock What determines the duration of stay of immigrants in {G}ermany?
  {E}vidence from a longitudinal duration analysis.
\newblock {\em International Journal of Social Economics\/}~{\em 35\/}(11),
  769--782.

\bibitem[\protect\citeauthoryear{Hirsch and Jahn}{Hirsch and
  Jahn}{2012}]{Hirsch&Jahn}
Hirsch, B. and E.~J. Jahn (2012).
\newblock Is there monopsonistic discrimination against immigrants? {F}irst
  evidence from linked employer-employee data.
\newblock IZA Discussion Papers 6472, Institute for the Study of Labor (IZA).

\bibitem[\protect\citeauthoryear{Hirsch, Schank, and Schnabel}{Hirsch
  et~al.}{2010}]{Hirschetal}
Hirsch, B., T.~Schank, and C.~Schnabel (2010).
\newblock Differences in labor supply to monopsonistic firms and the gender pay
  gap: {A}n empirical analysis using linked employer-employee data from
  {G}ermany.
\newblock {\em Journal of Labor Economics\/}~{\em 28\/}(2), 291--330.

\bibitem[\protect\citeauthoryear{Lehmer and Ludsteck}{Lehmer and
  Ludsteck}{2011}]{Lehmer&Ludsteck}
Lehmer, F. and J.~Ludsteck (2011).
\newblock The returns to job mobility and inter-regional migration: Evidence
  from germany.
\newblock {\em Papers in Regional Science\/}~{\em 90\/}(3), 549--571.

\bibitem[\protect\citeauthoryear{Manning}{Manning}{2003}]{manning2003book}
Manning, A. (2003).
\newblock {\em Monopsony in motion: imperfect competition in labor markets}.
\newblock Princeton University Press.

\bibitem[\protect\citeauthoryear{Mortensen and Neumann}{Mortensen and
  Neumann}{1988}]{Mortensen&Neumann}
Mortensen, D.~T. and G.~R. Neumann (1988).
\newblock Estimating structural models of unemployment and job duration.
\newblock In W.~A. Barnett, E.~R. Berndt, and H.~White (Eds.), {\em {D}ynamic
  {E}conometric {M}odeling: {P}roceedings of the third {I}nternational
  {S}ymposium in {E}conomic {T}heory and {E}conometrics}, pp.\  335--355.
  Cambridge University Press.

\bibitem[\protect\citeauthoryear{Sa}{Sa}{2011}]{Sa2011}
Sa, F. (2011).
\newblock Does employment protection help immigrants? {E}vidence from european
  labor markets.
\newblock {\em Labour Economics\/}~{\em 18\/}(5), 624--642.

\bibitem[\protect\citeauthoryear{van~den Berg}{van~den
  Berg}{2001}]{VanDenBerg2001}
van~den Berg, G.~J. (2001).
\newblock Duration models: {S}pecification, identification and multiple
  durations.
\newblock In J.~Heckman and E.~Leamer (Eds.), {\em Handbook of Econometrics},
  Volume~5, Chapter~55, pp.\  3381--3460. Elsevier.

\bibitem[\protect\citeauthoryear{van~den Berg and Ridder}{van~den Berg and
  Ridder}{1998}]{VanDenBerg&Ridder}
van~den Berg, G.~J. and G.~Ridder (1998).
\newblock An empirical equilibrium search model of the labor market.
\newblock {\em Econometrica\/}~{\em 66\/}(5), 1183--1222.

\bibitem[\protect\citeauthoryear{Velling}{Velling}{1995}]{Velling}
Velling, J. (1995).
\newblock Wage discrimination and occupational segregation of foreign male
  workers in {G}ermany.
\newblock ZEW Discussion Papers 95-04, Center for European Economic Research
  (ZEW).

\end{thebibliography}

\end{document}